\documentclass[10pt]{article}
\usepackage[utf8]{inputenc}
\usepackage{authblk}
\usepackage{dirtytalk, float, enumitem}
\usepackage[top=.8in, bottom=0.75in, left=1in, right=1in]{geometry}
\usepackage{epsfig, color}
\usepackage{amsfonts, amsmath, setspace}
\usepackage{mathtools, array, booktabs}
\usepackage{amssymb, cite, soul, url}
\usepackage[normalem]{ulem}
\usepackage{scalerel}
\usepackage{graphicx, float, caption, subcaption} 
\graphicspath{{Report_Figures/Figures_Wavelets/}} 
\usepackage{lineno}
\begin{document}

\title{Scaling analysis for buoyant plumes over wildland fires}
\author[1,2]{Ajinkya Desai}
\author[2]{Antonio Quim Cervantes}
\author[2,3]{Tirtha Banerjee}
\affil[1]{Lawrence Livermore National Laboratory\footnote{Release Number: LLNL-JRNL-2024559}, Livermore, CA 94550}
\affil[2]{Department of Civil and Environmental Engineering, University of California, Irvine, CA 92697}
\affil[3]{Department of Earth System Science, University of California, Irvine, CA 92697}\date{}
\maketitle

\section*{Abstract}
Tracking the structure and geometric properties of a buoyant plume in crosswind is critical for managing smoke hazards and improving disaster mitigation decision-making. Plume features, such as tilt angle, height, and curvature changes, are impacted by multiple forcing parameters, with canopy-induced turbulence patterns adding complexity. This study examines how these parameters, reduced to fewer dimensionless groups, affect the plume centerline slope, both near the surface and in the far field. Results from large-eddy simulations, conducted in both canopy and no-canopy environments, explore power-law dependencies, based on prior formulations, between the slopes and dimensionless groups describing the (1) relative strength of the buoyancy source to ambient wind forcing, (2) plume turbulence intensity relative to upstream turbulence, and (3) canopy density and height. Near-surface slopes are an order of magnitude higher in the canopy cases, although their sensitivity to group (1) is similar across both environments. In the canopy cases, the near-surface slope follows a one-fourth power-law dependence with group (2). This effect is absent in the no-canopy case, reflecting differences in the momentum-flux structure near the source.  Moreover, the canopy aerodynamic effects delay the plume transition from the rise phase into the far-field, bent-over phase; however, canopy effects diminish as buoyancy source strength increases. The near-surface plume slope exhibits a more complex dependence on canopy geometry, increasing almost linearly with the canopy height but showing weaker sensitivity to canopy drag. Our findings support the development of scaling laws for plume structures across varied vegetative landscapes and inform improved predictive modeling efforts.\\

\textbf{Keywords}: Buoyancy source strength, Canopy geometry, Plume centerline slopes, Power-law dependence, Turbulence intensity\\


\section{Introduction}

Quantifying the dynamics of plumes emanating from surface fires is crucial for assessing respiratory and visibility hazards due to smoke, heat transfer to flammable vegetation and property in the vicinity, and for informing fire management strategies. This includes an accurate, physics-based estimation of plume trajectories under the varying influencing conditions of fire intensity, ambient wind forcing, and geometry of the burning region \cite{himoto2020temperature}. Moreover, tall vegetative canopies significantly alter the coherent patterns associated with momentum transport and exchange with the atmosphere aloft, and turbulence intensities \cite{raupach1996coherent, finnigan2000turbulence, heilman2017atmospheric, heilman2021observations, desai2023features, desai2024investigating, desai2025investigating, banerjee2026role}. Therefore, trajectories of plumes with heat sources beneath a tall canopy can differ substantially from those in no-canopy environments.



A large volume of early literature is dedicated to laminar and turbulent buoyant plumes emerging from point or line sources in quiescent environments \cite{morton1956turbulent, morton1959forced, turner1969buoyant, hunt2011classical, van2010universal}, capitalizing on similarity criteria to develop scaling laws characterizing the spatially evolving plume structure. Insights from several recent studies have been aimed at strengthening the physical underpinnings and overcoming the perceived shortcomings of classical plume theory \cite{morton1956turbulent, morton1959forced} in relation to unsteady effects, corrections for source geometry, explosive thermals, etc.\cite{craske2016generalised, saeed2022buoyancy, webb2023turbulent, hogg2018unsteady, skvortsov2021scaling, kocan2022scaling}.
Additionally, over the last few decades, power-law scalings for buoyant plumes in cross flow have been developed using several experimental studies 
\cite{briggs1984plume, weil1988, davidson1994dimensionless, pun1999behaviour, robins2000water, lee2003} with the pioneering work by Briggs \cite{briggs1984plume} serving as the primary reference for point sources. An early study by Raupach \cite{raupach1990similarity} used dimensional analysis to model bushfire plumes as a line source of buoyancy. Based on a series of assumptions, including plume self-similarity under a cross-wind and extrapolation of point-source updraft velocity behavior to line sources, Raupach related plume angles to updraft velocity and buoyancy-source strength. Davidson \cite{davidson1994dimensionless} demonstrated that predictions of the plume trajectory from a rectified version of the two-thirds law for non-point (finite) sources, derived from conservation laws, showed improved agreement with flume experiments near the source and approached the two-thirds law downstream from the source (far-field region). Apart from validating a one-thirds law for plume rise in the near source region, with an added initial momentum at the source, and the two-thirds power law for plume rise in the far field regions, for both laminar and turbulent cross-flow, Huq and Stewart \cite{huq1996laboratory} demonstrated a lower plume rise in the presence of ambient turbulence using water tunnel experiments. In a subsequent analysis, based on LIF (laser-induced fluorescence) experiments on plume discharge from a finite source, Pun and Davidson  \cite{pun1999behaviour} provided evidence supporting a three-fourths law near the plume source, referred to as the \say{weakly-advected-plume region}, before undergoing a transition to the two-thirds law further away, called the \say{strongly-advected-thermal region}.  A recent study by James et al. \cite{james2022particle} corroborated the one-thirds and two-thirds laws using experiments conducted on single-phase saline plumes and line thermals in cross-flow, with coefficients in good agreement from those in previous studies \cite{hewett1971laboratory, chu1975turbulent}. Through wind-tunnel experiments that deployed rectangular diffusion burners, Himoto et al. \cite{himoto2020temperature} supplemented the literature by developing a power-law dependency of plume trajectories on dimensionless forms of not only the heat release rate and cross-wind velocity at the inlet, but also the the burner aspect ratio. The power-law model represented the experimental data to a high degree of accuracy, except close to the plume source. For propagating fires, Byram's energy criterion is used as a standard for determining whether a fire is buoyancy- or wind-dominated, with the plume tilt closely associated with Byram's convective number ($N_\text{c}$) \cite{gm1959combustion, nelson2015re, morandini2010experimental, zhang2020study}; however, an explicit power-law dependence between $N_\text{c}$ and the plume tilt has not been derived yet, particularly in forested environments.

Laboratory(-scale) studies are limited by their scale, which can impact the physics of the observed plume behavior. Computational modeling efforts have also been invested in exploring scaling laws and have served to draw comparisons with experiments. Devenish et al. \cite{devenish2010entrainment} compared outputs from large-eddy simulations (LES) with plume-trajectory power laws for buoyant plumes interacting with both weak and strong cross-flow under stable conditions. They found substantial evidence to support the two-thirds law for plume rise with downwind distance in strong cross-wind when analytically correcting for non-point (finite) source sizes, but little evidence for weak cross-wind. A recent LES study that incorporated similar conditions, by Cintolesi et al. \cite{cintolesi2019turbulent} and direct numerical simulations (DNS), by Jordan et al. \cite{jordan2022under}, on plumes, with zero velocity at the source and interacting with a uniform cross-wind, also corroborated the two-thirds law.


While buoyant plumes interacting with a cross-wind in open (no-canopy) environments have been the focus of most previous studies, with the effect of atmospheric stability gaining some attention \cite{dubey2024effect}, research on the added influence of vegetative canopies remains elusive. 
While Raupach \cite{raupach1990similarity} proposed scaling laws for different source geometries, the study primarily discussed the shearing effects of plant canopies qualitatively. More recently, Chung and Koseff \cite{chung2023interaction} proposed scaling parameters capturing the complex interplay between buoyancy, cross-wind, and canopy turbulence, which were based on the friction velocity and the Richardson number obtained from flume experiments with a canopy represented using an array of dowel rods. Field-scale experiments (e.g. controlled pile and broadcast burn experiments) that have studied turbulent plumes originating within tall vegetative canopies \cite{heilman2015observations, heilman2017atmospheric, heilman2019observations, desai2023features, desai2024investigating, banerjee2026role, heilman2023atmospheric, bebieva2020role}, though useful in their own right, are limited by the spatial localization of measurements and logistical obstacles to deploying extensive instrumentation while maintaining quality control, making the development of scaling laws for characterizing plume dispersion on field scales an uphill task. Resorting to LES, Desai et al. \cite{desai2025investigating} documented differences in the plume centerline deflection under cross-wind, both near the heat source and in the far field, between a canopy and no-canopy setting. It was found that the near-surface plume centerline deflection from the vertical was much weaker in the canopy case due to canopy drag and the presence of a leeward recirculation zone that obstructed the upstream flow. Additionally, the plume centerline depicted a three-part structure in the canopy case, with an intermediate region between the near-surface and far-field regions, which was absent in the no-canopy case. Moreover, the plume transitioned to the bent-over phase in the far-field at a higher altitude in the canopy setting. However, the effect of changing the buoyancy strength relative to ambient forcing and altering the canopy height and density were not studied. In an adjacent study, Cervantes et al. \cite{cervantes2026impact} compared canopy edge and gap flow effects on a buoyant plume in cross-wind versus homogeneous- and no-canopy surface conditions. However, deviations of the plume centerline geometry from those observed in the latter scenarios remain to be quantified. 

Dimensional analysis has found applications in several fluid mechanics problems and can help untangle the complex dynamics of turbulent flows in a variety of physical (boundary) conditions \cite{gibbings2011dimensional, katul2019primer}. In this case, it presents a powerful tool to condense multiple dominant factors governing plume deflection into fewer dimensionless groups. These groups can be used to develop scaling laws governing the dependence of plume trajectories on a range of parameters, including buoyancy source strength, ambient wind forcing, and heat-source geometry, as demonstrated by many previous studies \cite{turner1969buoyant, raupach1990similarity, himoto2020temperature}. For instance, a recent summary paper by Taherian and Mohammadian \cite{taherian2021buoyant} attempted a dimensional analysis for a buoyant jet in cross-flow and no-canopy conditions, relating trajectories to a cross-flow-based densimetric Froude number, without obtaining an explicit functional relation.



In this paper, we propose scaling relationships characterizing plume rise and dispersion with a particular emphasis on canopy environments. This is accomplished via a dimensional analysis (Sect.~\ref{sect_methods}) followed by an analytical investigation into the consistency of power-law scaling dependencies on the dimensionless groups based on key studies (Sect.~\ref{sect_liter_scaling}), and a characterization of plume slopes based on a large suite of large-eddy simulations comparing buoyant plumes in canopy and \say{no-canopy} environments (Sect.~\ref{sect_les_scaling}). Through such comparisons, our analysis provides insights into the relative importance of the underlying physical mechanisms informing buoyant plume trajectories, both near the surface and in the far-field, when the surface boundary conditions reflect canopy presence (Sect.~\ref{sect_conclusions}). Scaling laws, based on the power-law dependencies explored in our analysis, can significantly reduce the computational overhead in field-scale fire-plume and smoke behavior models \cite{himoto2020temperature}. The analysis informs the development of improved predictive models for plume rise and smoke transport in a range of conditions, which is important for managing air quality hazards during natural and human-made disasters and when planning prescribed burns for preemptive reduction of hazardous fuels. 


\section{Methods}
\label{sect_methods}
Starting from a summary of the primary physical variables involved, which govern plume deflection, we first obtain our dimensionless groups based on the Buckingham Pi Theorem (BPT) \cite{gibbings2011dimensional, katul2019primer}. We then revisit scaling laws governing the plume trajectory from some representative studies in both canopy and open environments and show that the plume centerline slope can be obtained in terms of power laws involving our dimensionless ($\pi$) groups despite differences in power-law exponents among various studies. Finally, we utilize outputs from a corpus of large-eddy simulations in a tall-canopy environment to explore a power-law dependency between the plume slope and other dimensionless parameters. These are complemented by a similar number of simulations in a no-canopy environment, with an otherwise similar computational domain, surface heat flux conditions, heat-source geometry, and wind-forcing conditions, to facilitate a comparison.

\begin{figure}[h!]
    \centering
    \includegraphics[scale=0.2]{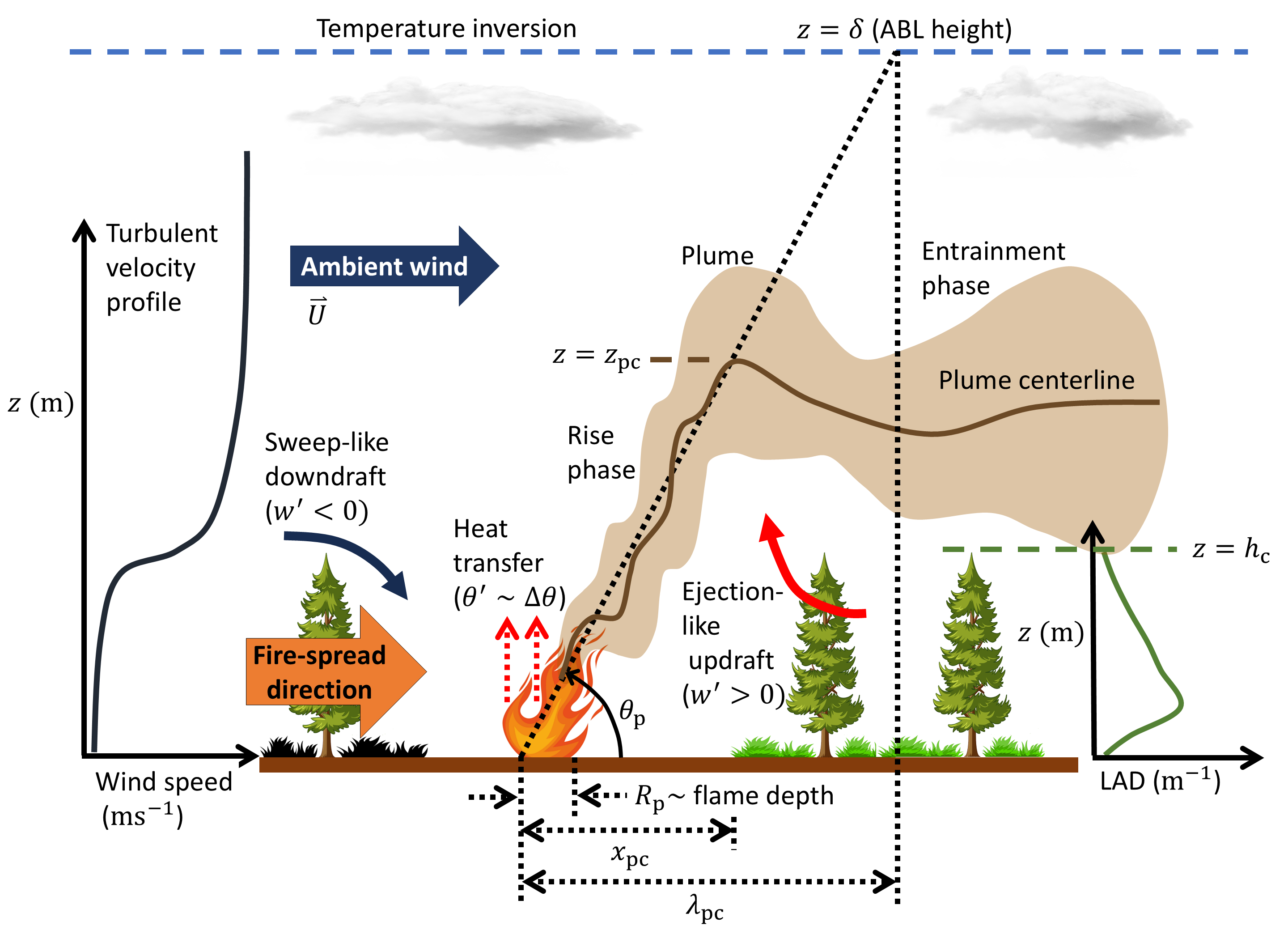}
    \caption{Schematic 2-D representation of the plume interacting with a cross-wind based on a linear plume rise model. After an initial rise phase, plume centerline trajectory transitions to the bent-over phase (sometimes called the entrainment phase) at $(x_\text{pc},~z_\text{pc})$, where the centerline undergoes a change in slope}
    \label{fig_schematic}
\end{figure}

Figure~\ref{fig_schematic} presents a simplified, 2-D picture of a plume, arising from a surface heat source, interacting with a cross-wind, the plume centerline trajectory transitioning from the rise phase to the far-field bent-over (or entrainment) phase at $(x_\text{pc},~z_\text{pc})$, along with a summary of the physical variables/parameters involved. It can be seen that the plume centerline slope that we aim to track in the analysis to follow is given by $z_\text{pc}/x_\text{pc}$, which is equivalent to $\lambda_\text{pc}/\delta$ by basic proportionality arguments. The variable $\lambda_\text{pc}$ is the horizontal length scale representing the onset of the bent-over phase. One way of interpreting $\lambda_\text{pc}$ is the streamwise length traversed by the plume, should the plume transition from the rise phase to the bent-over phase at the height of the inversion layer, i.e. if $z_\text{pc}\rightarrow\delta$.

\subsection{Obtaining Dimensionless ($\pi$) Groups}
The key variables and their dimensional representations are summarized in Table~\ref{tab_var}. 
\begin{table}[h!]
    \centering
    \begin{tabular}{p{5mm}|p{22mm}|p{75mm}|p{15mm}|p{18mm}}
    \hline
\textbf{No.} &   \textbf{Category} &  \textbf{Variable} & \textbf{Symbol} & \textbf{Dimensions} \\
    \hline
1 &    Response &   Horizontal length scale, plume bend & $\lambda_\text{pc}$ & $L$\\
2&   Geometric constraint & Atmospheric Boundary Layer (ABL) height & $\delta$ & $L$ \\
  
3 &  Forcing &  Buoyancy & $g$ & $L T^{-2}$ \\

4& Forcing &   Kinematic heat flux difference at surface ($\Delta H_\text{s} = H_\text{patch}-H_\text{surface}$) & $\frac{\Delta H_\text{s}}{\rho c_\text{p}}$ & $\Theta L T^{-1}$ \\
5& Forcing &   Friction velocity (turbulence intensity) & $u_*$ & $LT^{-1}$ \\

6& Boundary condition & Canopy height & $h_\text{c}$ & $L$ \\
7& Geometric constraint &  Canopy-drag (adjustment) length scale & $L_\text{c}$ & $L$ \\
8&  Boundary condition &  Effective radius of surface heat source patch (Heat source area / Perimeter) & $R_\text{p}$ & $L$ \\

9& Forcing &   Ambient horizontal velocity & $U$ & $LT^{-1}$ \\
10& Forcing & Ambient potential temperature & $\theta_\text{a}$ & $\Theta$ \\
    \hline
    
\end{tabular}
    \caption{A summary of the variables used in the dimensional analysis}
\label{tab_var}
\end{table}
 Note that we have ignored two variables, i.e. kinematic viscosity ($\nu$) and surface roughness ($z_0$), since the surface roughness is much shorter compared to other relevant length scales (it can be argued that we are considering the average flow outside of the influence of the surface roughness) and the flow is at a sufficiently high Reynolds number to preclude viscous effects. We begin with a functional dependence of the response variable $\lambda_\text{pc}$ to the geometric constraints, forcing variables, and boundary conditions of the following form: $\lambda_\text{pc} = \mathcal{F}(\delta,~g, \frac{\Delta H_\text{s}}{\rho c_\text{p}},~u_*,~h_\text{c},~L_\text{c},~R_\text{p},~U,~\theta_\text{a})$. Since we have 10 variables ($M_\text{d}=10$) and 3 fundamental dimensions ($N_\text{d}=3$), i.e. $ L, T, \Theta$, the number of dimensionless groups is: $M_\text{d}-N_\text{d}= 10 - 3 = 7$. 
We choose the repeating variables such that they collectively involve the three fundamental dimensions: (i) $R_\text{p}$ [$L$], (ii) $U$ [$L,~T$], and (iii)  $\theta_\text{a}$ [$\Theta$]. 
These repeating variables are chosen because they are relatively easy to control and can be readily determined from numerical models of fire behavior or field experiments. The formation of the $\pi$ groups, based on the BPT, is presented in the Appendix for brevity and yields the following expressions. \\

\noindent\textbf{In the canopy case:} $\pi_1 = f(\pi_2, \pi_3, \cdots, \pi_6, \pi_7 )$
\begin{equation}
\begin{split}
\pi_1 &= ~~f(\pi_2, ~~~~\pi_3, ~~~~\pi_4, ~~~~~~~\pi_5, ~~\pi_6, ~~ \pi_7),\\
  \frac{\lambda_\text{pc}}{R_\text{p}}   &= f\left(\frac{\delta}{R_\text{p}},~g\frac{R_\text{p}}{U^2}, ~\frac{\Delta H_\text{s}}{\rho c_\text{p}}\frac{1}{\theta_\text{a}U}, \frac{u_*}{U}, \frac{h_\text{c}}{R_\text{p}}, \frac{L_\text{c}}{R_\text{p}}\right),
  \end{split}
  \label{eq_pi_canopy}
\end{equation}
where $ L_\text{c} = (c_\text{d}LAD)^{-1}$ and is referred to as the canopy-drag or adjustment length scale \cite{katul2013scalar}. Since the integral of the leaf-area density ($LAD$) over height is the leaf-area index ($LAI$), i.e. $\int_{0}^{h_\text{c}}LAD\,\text{d}z = LAI$, the adjustment length scale can be approximated using the $LAI$, i.e. $L_\text{c} \approx h_\text{c}(c_\text{d}LAI)^{-1} $.  \\

\noindent\textbf{In the no-canopy case, there are only five $\pi$ groups, i.e. $\pi_1 = f(\pi_2, \pi_3, \pi_4, \pi_5)$:}
\begin{equation}
\begin{split}
\pi_1 &= ~~f(\pi_2, ~~~~\pi_3, ~~~~\pi_4, ~~~~~~~\pi_5),\\
  \frac{\lambda_\text{pc}}{R_\text{p}}   &= f\left(\frac{\delta}{R_\text{p}},~g\frac{R_\text{p}}{U^2}, ~\frac{\Delta H_\text{s}}{\rho c_\text{p}}\frac{1}{\theta_\text{a}U}, \frac{u_*}{U}\right),
 \end{split}
  \label{eq_pi_nocanopy}
\end{equation}
where the terms associated with canopy presence and characteristics are precluded. 
According to the linear plume rise model (Fig.~\ref{fig_schematic}), the plume slope in the rise phase for both cases is recovered as follows: 

\begin{equation*}
    \frac{\delta}{\lambda_\text{pc}} = \frac{\delta}{R_\text{p}}\frac{1}{f}.
\end{equation*}

\noindent In addition, $\pi_4$ can be recast as follows:

\begin{equation*}
    \frac{\Delta H_\text{s}}{\rho c_\text{p}}\frac{1}{\theta_\text{a}U} \simeq \frac{\overline{w'\theta'}}{u_*}\frac{1}{\theta_\text{a}}\frac{u_*}{U}.
\end{equation*}

\noindent We know that $u_*^2 \simeq |u'w'|$. If we reasonably assume that $u'\sim u_*$, then $w'\sim u_*$. Moreover, we can argue that $\theta'\sim \Delta\theta = \theta - \theta_\text{a}$, the equation above becomes: 
\begin{equation*}
    \frac{\Delta H_\text{s}}{\rho c_\text{p}}\frac{1}{\theta_\text{a}U} \sim \frac{\Delta\theta}{\theta_\text{a}}\frac{u_*}{U}\implies \pi_4 \sim \frac{\Delta\theta}{\theta_\text{a}}\pi_5.
\end{equation*}

\noindent Moreover, we note the following. For an ideal gas, at atmospheric pressure in the environment, $ p_\text{a} \approx \rho_\text{a} R \theta_\text{a}$, where the absolute temperature can be approximated to be the potential temperature near the surface. Under thermal expansion, 
\begin{align*}
    p_\text{a} = (\rho_\text{a} + \Delta\rho) R (\theta_\text{a} + \Delta\theta) \implies p_\text{a} = \rho_\text{a} R\theta_\text{a} \left(1+\frac{\Delta\rho}{\rho_\text{a}}\right)\left(1+\frac{\Delta\theta}{\theta_\text{a}}\right),\\
    \implies 1 = 1 + \frac{\Delta\rho}{\rho_\text{a}} + \frac{\Delta\theta}{\theta_\text{a}} + \text{h.o.t.}
    \implies\frac{\Delta\rho}{\rho_\text{a}} = - \frac{\Delta\theta}{\theta_\text{a}}.     
\end{align*}
Additionally, by definition of the coefficient of thermal expansion ($\beta_\text{T}$), 
\begin{align*}
    \Delta\rho = - \beta_\text{T}\rho_\text{a}\Delta\theta \implies
     \beta_\text{T}\Delta\theta = -\frac{\Delta\rho}{\rho_\text{a}} = \frac{\Delta\theta}{\theta_\text{a}}.
\end{align*}
Therefore, 
\begin{equation}
\frac{\pi_4}{\pi_5} = \frac{\Delta H_\text{s}}{\rho c_\text{p}}\frac{1}{\theta_\text{a}u_*} \sim \frac{\Delta\theta}{\theta_\text{a}} \equiv -\frac{\Delta\rho}  {\rho_\text{a}} \equiv \beta_\text{T}\Delta\theta.
\label{eq_pi4pi5}
\end{equation}

\subsection{Computational Setup and Parameters}
\label{sect_setup}
We use the Parallelized Large-Eddy 
Simulation Model (PALM) for the LES, which solves the non-hydrostatic, incompressible Navier-Stoke equations, filtered by the grid size, in Boussinesq-approximated form \cite{maronga2020overview}. The computational setup reflects that utilized by Desai et al. \cite{desai2025investigating}. The computational domain is 2\,km long in the streamwise ($x$) direction, 1\,km wide in the cross-stream ($y$) direction, and 1.1\,km in the vertical direction ($z$). The cell size for the structured grid is 5\,m$ \times$5\,m$ \times$2\,m up to $z=2h_\text{c}$, $h_\text{c}$ being the canopy height, beyond which vertical grid stretching is employed. The vertical grid spacing ($\Delta_z$) increases in geometric multiples of the stretch factor (1.08), until the maximum cell size of 22\,m in the vertical direction is achieved at $z\approx340$\,m. The higher resolution near the surface is intended to capture the complex turbulence patterns in the canopy sub-layer. 
Fire plume presence is simulated by a
thermal plume arising from a localized region of area 100\,m\,$\times$\,100\,m with a prescribed high sensible heat flux. For the base case simulated by Desai et al. \cite{desai2025investigating}, the surface heat flux in this region was taken to be 100 times the ambient surface sensible heat flux  ($H_\text{surface} = 50$\,Wm\textsuperscript{-2}, $H_\text{patch}=5$\,kWm\textsuperscript{-2}), with a geostrophic wind speed of 2\,m\,s\textsuperscript{-1}. For the parametric study in this analysis, the heat flux at the source is increased in steps of  5\,kWm\textsuperscript{-2} in the range 5--35\,kWm\textsuperscript{-2}, while the geostrophic wind speeds ($U$) considered include 2, 3, 5, and 7\,m\,s\textsuperscript{-1}. Together, for the canopy and no-canopy environments, outputs from a total of 54 simulations are used. The ambient wind follows a turbulent, logarithmic profile at the inlet, which morphs into a profile with an inflection point near the canopy top over some fetch. Daytime atmospheric conditions are simulated with a well-mixed layer, i.e. a uniform initial profile of the potential temperature with height, up to a temperature inversion height ($\delta$) of approximately 1.1\,km. A more comprehensive description of the LES model and computational setup can be obtained from Maronga et al. \cite{maronga2020overview} and Desai et al. \cite{desai2025investigating}, with Cervantes et al. \cite{cervantes2026impact} validating the setup against results from experiments in no-fire conditions. The effect of model constraints such as the domain size and fetch on the plume physics in the canopy environment has been shown to be minimal for the base case (Desai et al, manuscript in press, \textit{Journal of Geophysical Research: Atmosphere}).

We rely on PALM's in-built plant-canopy module to simulate the presence of a forest canopy at the surface. The canopy elements are not explicitly modeled; rather, a volume-averaged effect of the porous canopy layer is considered, with the canopy drag-force term in the momentum equation for the mean velocity accounting for the aerodynamic effects of the canopy. This term is given by $f_{\text{D}_{i}} = -c_\text{d}LADu_i\sqrt{u_{i}^{2}}$, where $u_i$ represents velocity components in the index notation and $LAD$ is the leaf-area density. Note that this implicitly assumes uniformity in the leaf area exposed to the flow in both the horizontal and vertical directions. To investigate the dependence of the plume deflection on the ambient wind speeds and surface sensible heat flux of the heat source as described above, a homogeneous canopy is simulated with $h_\text{c}=$35\,m, a drag coefficient ($c_\text{d}$) of 0.2, and a leaf area index ($LAI$) of 5 with the corresponding $LAD$ profile taken from Dias-Junior et al. \cite{dias2015large}. In addition, to examine the complex impacts of canopy height and drag characteristics, a suite of simulations was conducted with $LAI =2,~3, [\cdots]~ 7$, 25\,m$~\leq h_\text{c}\leq 51$\,m and $0.15\leq c_\text{d}\leq 0.4$ to construct the parametric space. These values are based on previous documentation on realistic canopy characteristics \cite{katul2004one, dias2015large, markkanen2003footprints, chaudhari2026impact, tota2012amazon}. The $LAD$ profile in each simulation is based on a Beta distribution \cite{markkanen2003footprints}, scaled by the $LAI$ and canopy height:

\begin{equation}
    LAD\left(z, a, b\right) = \frac{\left(\frac{z}{h_\text{c}}\right)^{a-1}\left(1-\frac{z}{h_\text{c}}\right)^{b-1}}{\int_0^1 \left(\frac{z}{h_\text{c}}\right)^{a-1}\left(1-\frac{z}{h_\text{c}}\right)^{b-1}\text{d}\left(\frac{z}{h_\text{c}}\right)} \times \frac{LAI}{h_\text{c}};~a, b > 1.
\end{equation}

\begin{figure}
    \centering
    \begin{tabular}{ccc}
       \includegraphics[width=0.3\linewidth, trim={3mm 1mm 5mm 3mm}, clip]{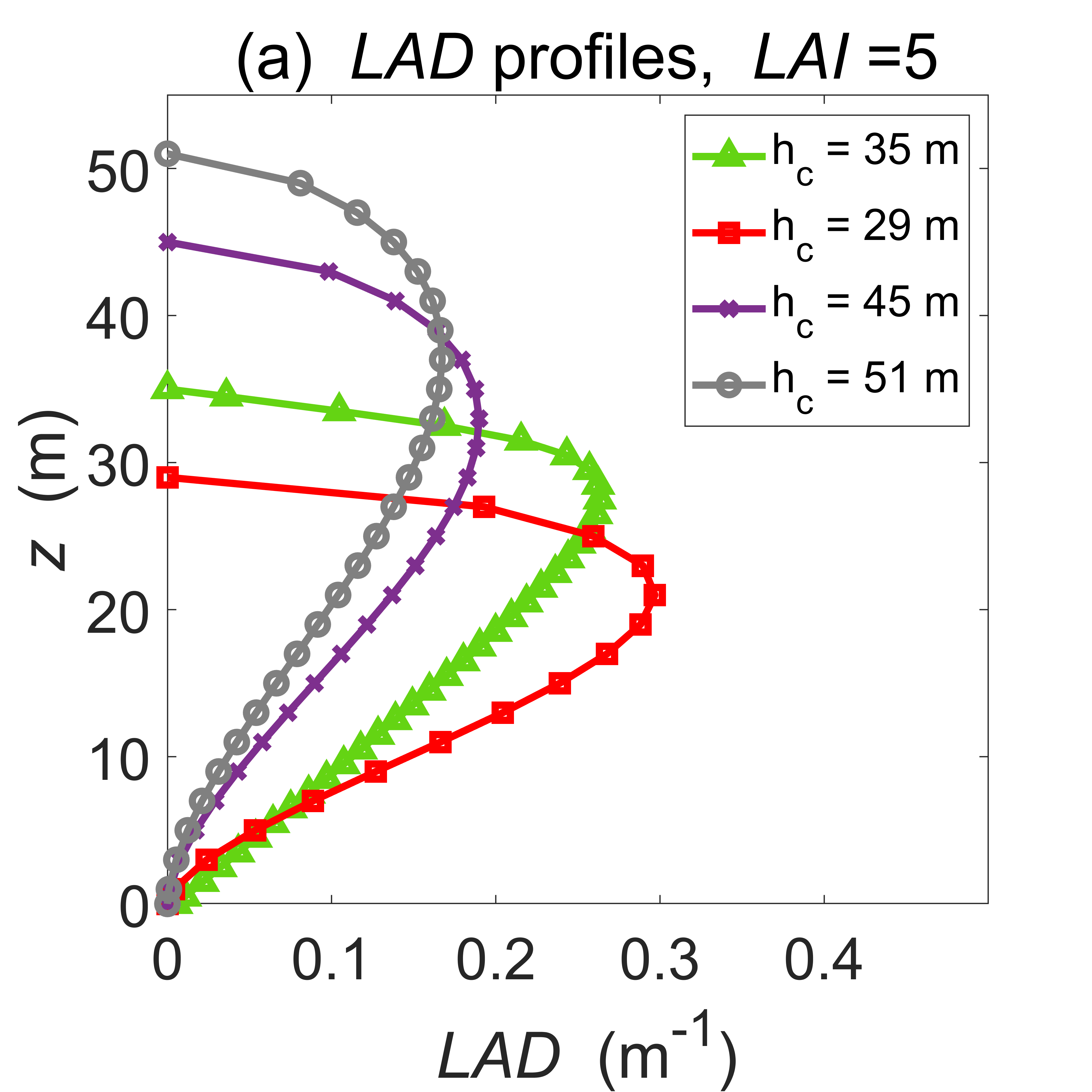} & \includegraphics[width=0.3\linewidth, trim={3mm 1mm 5mm 3mm}, clip]{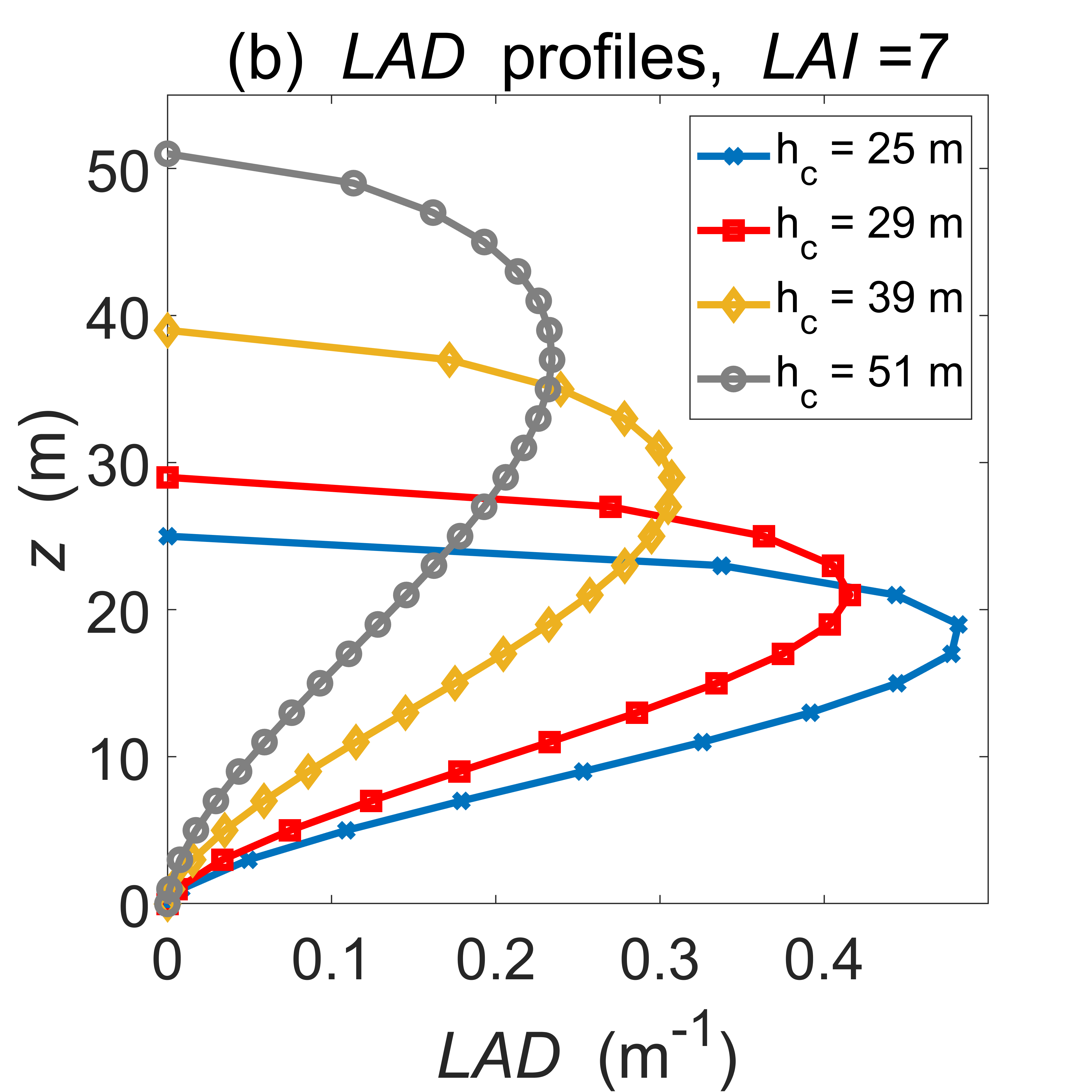} & \includegraphics[width=0.302\linewidth, trim={3mm 1mm 5mm 3mm}, clip]{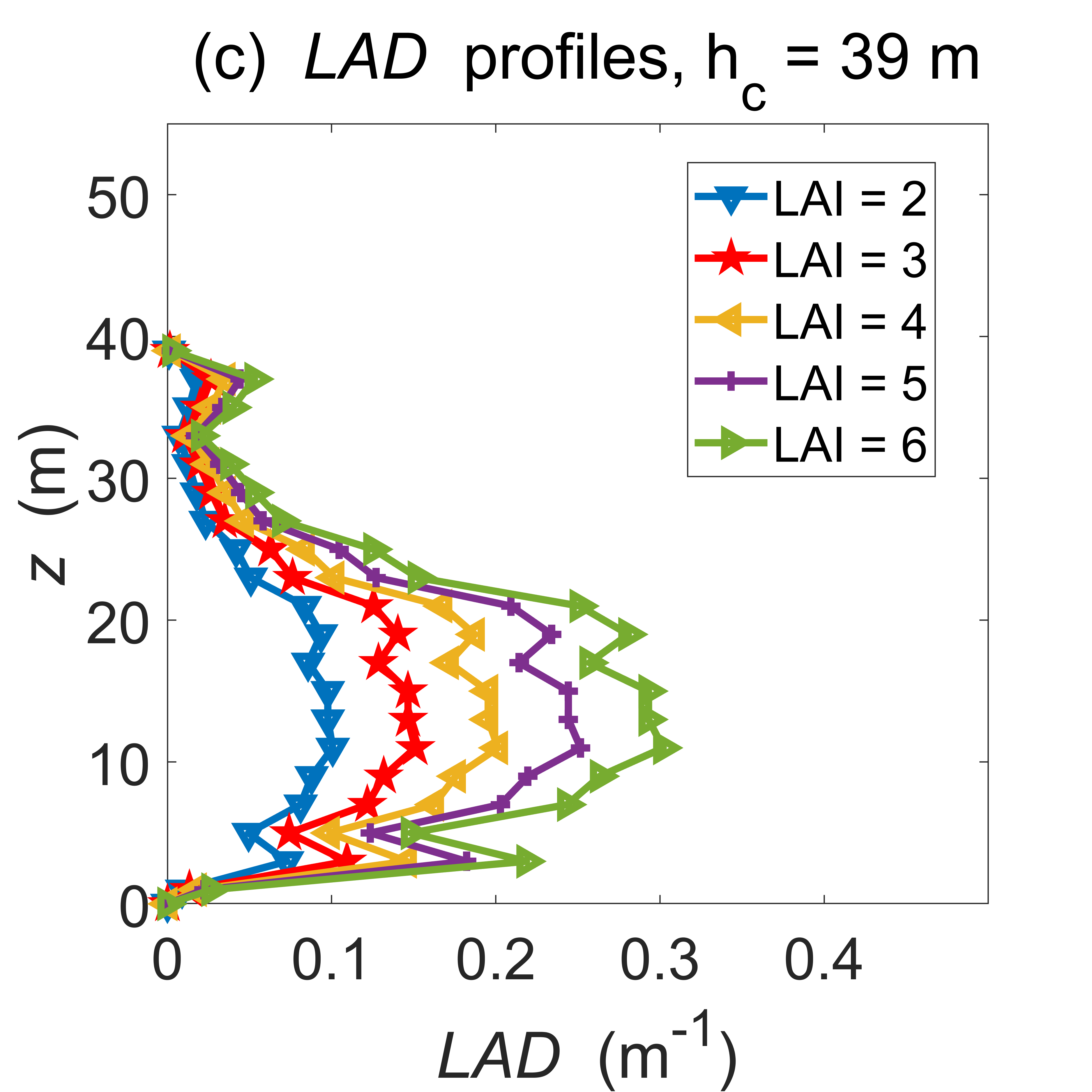}
    \end{tabular}
   
    \caption{Selected examples of $LAD$ profiles used in the canopy simulations for (a) $LAI=5$, with $h_\text{c}=5$ representing the digitized profile from Dias-Junior et al. \cite{dias2015large}; for (b) $LAI=7$; (c) all bottom-heavy profiles included in this study for $h_\text{c}=39$\,m based on a scaled profile from Tota et al. \cite{tota2012amazon}
    }
    \label{fig_lad_profiles}
\end{figure}

\noindent Here, the parameters $a$ and $b$, as determined by fitting a Beta distribution to the $LAD$ profile of the canopy simulated by Dias-Junior et al. \cite{dias2015large}, using a simplex algorithm (MATLAB's \textit{fminsearch}() function), i.e. $a=2.67, b=1.61$, are utilized. Note that this results in a top-heavy tree canopy, as shown for some examples of $LAI$ and $h_\text{c}$ in Figs.~\ref{fig_lad_profiles}(a)--(b). In addition, we have included simulations with realistic $LAD$ profiles for $h_\text{c}=39$\,m; these are obtained from a digitized version of the $LAD$ profile for an Amazon rainforest with $LAI=7$ \cite{tota2012amazon} scaled by $LAI=2,~3,[\cdots],~6$, and are intended to diversify the simulations for our scaling analysis (Fig.~\ref{fig_lad_profiles}(c)). Overall, results from 36 simulations are used to explore the dependence on canopy geometry with 8.9\,m$~\leq L_\text{c}\leq127.5$\,m. In addition, we include plume deflection results from four configurations investigating the effects of canopy heterogeneity for two wind speeds each, i.e. $U=5,~10$\,m\,s\textsuperscript{-1}, studied by Cervantes et al. \cite{cervantes2026impact}. These include the surface heat source ($100$\,m$\times100$\,m) positioned in a 100-m gap in an otherwise continuous canopy (G100), in a 200-m gap (G200), immediately on the outside of the upstream canopy edge (EE), and immediately on the inside of the upstream edge (IE). More details about the setup for the heterogeneous-canopy cases are provided by Cervantes et al. \cite{cervantes2026impact}.

\section{Results}
\subsection{Consistency with scaling laws from prior studies}\label{sect_liter_scaling}

We show that the $\pi$ groups obtained above are consistent with scaling laws for stationary sources of different geometries as obtained in previous literature. We start with power-law scalings for plume trajectories outlined in certain representative studies on buoyant plumes in cross-flow, most of which preclude the presence of a canopy, given the larger body of work under those conditions as opposed to canopy presence. Next, through mechanistic and scaling arguments, we obtain power-law scalings for the plume centerline slope as functions of the $\pi$ groups in Eqs.~\eqref{eq_pi_canopy} and~\eqref{eq_pi_nocanopy}.

  \subsubsection{Line source, cross-wind, no canopy}\label{sect_line_nocan}
Raupach \cite{raupach1990similarity} obtained the following similarity law for buoyant plumes in cross-wind, purely from dimensional considerations: 
\begin{align*}
\frac{\overline{z}}{x} \sim \frac{B^{1/3}}{u}, \text{where}~B = \int_{-\infty}^{+\infty} w g\frac{\Delta \theta}{\theta_\text{a}}\text{d}x~\text{and represents the buoyant flux per unit length}.
\end{align*}
Additionally, they argued that the mean slope of the plume centerline in the rise phase is given by the ratio of the vertical velocity to the horizontal: 
\begin{align*}
    \frac{\text{d}\overline{z}}{\text{d}x} = \frac{w}{u}.
\end{align*}
While this is a reasonable method to estimate the plume centerline slope in the rise phase and has been adopted by many studies \cite{chung2023interaction}, it is based on the assumption that the mean plume centerline trajectory follows the average streamline within the plume, which has been challenged in the literature \cite{jordan2022under}. We relax this assumption by replacing the equality above with a similitude. Moreover, we perceive $w$ and $u$ as vertical and horizontal velocity scales, rather than local components of velocity, i.e. $u \sim U$. For a plume model with a linear rise phase, as shown in Fig.~\ref{fig_schematic}, we get: 
\begin{align*}
  \frac{w}{u} \sim  \frac{\text{d}\overline{z}}{\text{d}x} \simeq \frac{z_\text{pc}}{x_\text{pc}} \sim  \frac{B^{1/3}}{u} \implies \frac{w^3}{u^3} \sim \frac{B}{u^3}.
\end{align*}
It follows that $w^3 \sim B$, which is precisely the vertical velocity scaling obtained by Raupach \cite{raupach1990similarity} for quiescent conditions from dimensional considerations. 
Thereby, Raupach's analysis assumes that the presence of a cross-wind merely alters the plume slope without changing the centerline vertical velocity behavior with height, or with streamwise distance from the source, from that observed in quiescent conditions. We will return to this assumption later in our analysis. Now, using Raupach's definition of $B$ for a line source:
    \begin{align}
    B = \int_{-\infty}^{+\infty} w g\frac{\Delta \theta}{\theta_\text{a}}\text{d}x \sim w g\frac{\Delta \theta}{\theta_\text{a}}L_\text{x}.
    \label{eq_buoyancy_line}
    \end{align}
    Here, $L_\text{x}$ represents the equivalent width of a line (high-aspect-ratio) heat source. \text{Then, noting that } $u \sim U~\text{and, using}~L_\text{x}\sim  R_\text{p}$ and Eq.~\eqref{eq_pi4pi5}, we get:    
    \begin{align*}
    {w^3} &\sim w g\frac{\Delta \theta}{\theta_\text{a}}L_\text{x} \Rightarrow
     \frac{w^2}{u^2} \sim \frac{g L_\text{x}}{u^2}\frac{\Delta\theta}{\theta_\text{a}} \Rightarrow \frac{w}{u} \simeq \sqrt{\frac{g R_\text{p}}{U^2}\frac{\Delta\theta}{\theta_\text{a}}}\sim\sqrt{\frac{\pi_3 \pi_4}{\pi_5}}.
\end{align*}
\text{Therefore, the plume centerline slope is given by:} 
\begin{equation}
   \frac{\delta}{\lambda_\text{pc}} \sim \sqrt{\frac{\pi_3 \pi_4}{\pi_5}}.
 \label{eq_slope_line}
\end{equation}

\noindent We note here that Raupach's assumption of $w^3 \sim B$ has also been used in subsequent studies to demonstrate the equivalence of the Clark convective Froude number ($F_\text{c}$) \cite{clark1996coupled1, clark1996coupled2, sullivan2007convective} and Byram’s energy criterion \cite{gm1959combustion, morandini2010experimental, nelson2015re}. 
Now, we use the definition of Byram's convective number ($N_\text{c}$) in terms of the fireline intensity ($I$) and fireline rate of spread ($r$) as follows, and note that $\text{for a static source},~r\rightarrow0$:
\begin{align*}
\lim_{r\rightarrow0}
    N_\text{c} = \lim_{r\rightarrow0} \left[2\frac{gI}{\rho c_\text{p}\theta_\text{a}(u-r)^3}\right] \implies N_\text{c} \sim \frac{gI}{\rho c_\text{p}\theta_\text{a}u^3}, ~\text{where}~B = \frac{gI}{\rho c_\text{p}\theta_\text{a}}.
\end{align*}
By Nelson's formula \cite{sullivan2007convective}, $I = \Dot{m}c_\text{p}\Delta\theta = \rho D w c_\text{p}\Delta\theta$, where $\Dot{m}$ is the mass flux rate and $D$ is the flame depth. Using these relations along with Eq.~\eqref{eq_slope_line}, we get:
\begin{align*}
    N_\text{c} \sim \frac{gD \Delta\theta}{ \theta_\text{a}u^3} w \sim \left[\frac{gD}{U^2}\right]\left[ \frac{\Delta\theta}{ \theta_\text{a}}\right]
    \frac{w}{u} \equiv \pi_3\frac{\pi_4}{\pi_5}\frac{\delta}{\lambda_\text{pc}},\text{~where~} D\equiv L_\text{x} \equiv R_\text{p}.
\end{align*}
Using  Raupach's assumption of $w^3 \sim B$, Sullivan \cite{sullivan2007convective} obtained the relation $N_\text{c} \sim F_\text{c}^{-3}$. If we use the definition of $F_\text{c}$ from Sullivan et al. \cite{sullivan2007convective}, for a static source ($r\rightarrow0$), and recast it in terms of our $\pi$ groups, we get:
\begin{align*}
    F_\text{c}^{-3} \sim \left( \frac{gD}{u^2}\frac{\Delta\theta}{\theta}\right)^{3/2} \equiv [\pi_3\pi_4\pi_5^{-1} ]^{3/2}.
\end{align*}
This gives us $N_\text{c} \sim \left[\sqrt{\pi_3\pi_4\pi_5^{-1}}\right]^{3}$, which helps us recover $\frac{\delta}{\lambda_\text{pc}} \sim \sqrt{\pi_3 \pi_4\pi_5^{-1}}$ as obtained in Eq.~\eqref{eq_slope_line}. Moreover, Byram's convective number varies as the cube of the plume centerline slope in the rise phase, i.e. $N_\text{c}\sim \left(\frac{\delta}{\lambda_\text{pc}}\right)^3$.  In this manner, we have shown that the plume centerline slope and Byram's convective number for a line source of buoyancy vary as functions of our $\pi$ groups.


\subsubsection{Point and finite sources, cross-wind, no-canopy: far-field}
Davidson \cite{davidson1994dimensionless} provides the expression for the plume centerline trajectory for a finite source with equivalent radius $R_\text{s}$ as follows, provided that the origin for $(x,~z)$ lies at a virtual point source to which the finite source can be extrapolated: 
\begin{equation}
    \frac{\overline{z}}{L_\text{B}} \sim \left(\frac{x}{L_\text{B}}\right)^{2/3},~\text{where}~L_\text{B} = g\frac{w_\text{s}R_\text{s}^2}{U^3}\frac{\Delta\theta}{\theta_\text{a}}.
    \label{eq_traj_point_far}
\end{equation}
Here, $w_\text{s}$ is the vertical velocity near the source and we have replaced $\Delta\rho/\rho_\text{a}$ in the original expression with $\Delta\theta/\theta_\text{a}$. The analysis corroborates similar expressions obtained from other studies \cite{davidson1994dimensionless, pun1999behaviour, lee2003, himoto2020temperature} for $\overline{z}>>L_\text{B}$, which is more likely to occur in the rise phase when the buoyancy is \say{weaker} compared to the ambient wind forcing. A similar expression for the plume centerline trajectory in the bent-over phase was previously obtained by Raupach \cite{raupach1990similarity} for a point source in cross-wind: 
\begin{align*}
    \frac{\overline{z}}{z_{U_0}} \sim \left(\frac{x}{z_{U_0}} \right)^{2/3},~\text{where}~z_{U_0} = \frac{B}{U^3},~~B = \int_{-\infty}^{+\infty} w g\frac{\Delta \theta}{\theta_\text{a}}2\pi r\text{d}r \equiv g\frac{\Delta \theta}{\theta_\text{a}} w_\text{s} R_\text{s}^2. 
\end{align*}
Here, we have expressed the buoyancy source strength using some equivalent radius $R_\text{s}$ and vertical velocity $w_\text{s}$. 
It follows that $z_{U_0} \equiv L_\text{B}$. In other words, a point source is interchangeable with a finite source of the same buoyancy source strength vis-\'a-vis the two-thirds law for $\overline{z}>>L_\text{B}$. Next, using Eq.~\eqref{eq_traj_point_far},  we have
\begin{equation}
    \frac{\overline{z}^3}{L_\text{B}^{3}} \sim \frac{x^2}{L_\text{B}^{2}} \Rightarrow \overline{z}\left(\frac{\overline{z}}{x}\right)^2 \sim L_\text{B}.
    \label{eq_nocan_far_transht}
    \end{equation}
Setting $x\rightarrow\lambda_\text{pc}$ as $\overline{z}\rightarrow\delta$, choosing $R_\text{p} = R_\text{s}$, and using $w_\text{s}/U \equiv w/U \simeq \delta/\lambda_\text{pc}$ and Eq.~\eqref{eq_pi4pi5}, we get
    \begin{align}
    \left(\frac{\delta}{\lambda_\text{pc}}\right)^2 \sim \frac{1}{\delta}\frac{g\Delta\theta}{\theta_\text{a}}\frac{w_\text{s}}{U^3}R_\text{s}^2 \equiv \frac{gR_\text{s}}{U^2}\frac{\Delta\theta}{\theta_\text{a}}\frac{w_\text{s}}{U}\frac{R_\text{s}}{\delta} \implies\frac{\delta}{\lambda_\text{pc}} \sim \frac{gR_\text{p}}{U^2}\frac{\Delta\theta}{\theta_\text{a}}\frac{R_\text{p}}{\delta}  \implies \frac{\delta}{\lambda_\text{pc}} \sim \frac{\pi_3\pi_4}{\pi_5\pi_2}.
    \label{eq_slope_point_far}\end{align}
    
\noindent Again, Eq.~\eqref{eq_slope_point_far} demonstrates the functional dependence of the plume centerline slope on our $\pi$ groups.

\subsubsection{Point and finite sources, cross-wind, no canopy: near-field}
Lee and Chu \cite{lee2003} suggested the following power-law scaling close to the plume source, i.e. for $\overline{z}<<L_\text{B}$, based on previous experiments:
\begin{equation*}
    \frac{\overline{z}}{L_\text{B}} \sim \left(\frac{x}{L_\text{B}}\right)^{3/4}
\end{equation*}
In situations where the ambient cross-wind is \say{weak} compared to the buoyancy forcing, it is possible to have a longer buoyancy length scale $L_\text{B}$. Under such conditions, if the transition height $z_\text{pc}$, as shown in Fig.~\ref{fig_schematic}, is shorter than the buoyancy length scale, i.e. $z_\text{pc}<<L_\text{B}$, the three-fourth law above may be applicable. In such a case, 
\begin{align}
    \frac{\overline{z}}{L_\text{B}} &\sim \left(\frac{x}{L_\text{B}}\right)^{3/4}\Rightarrow \left(\frac{\overline{z}}{x}\right)^{3} \sim \frac{L_\text{B}}{\overline{z}} 
    \label{eq_nocan_near_transht}
    \end{align}
    Setting $x\rightarrow\lambda_\text{pc}$ as $\overline{z}\rightarrow\delta$, choosing $R_\text{p} = R_\text{s}$, and using $w_\text{s}/U \equiv w/U \simeq \delta/\lambda_\text{pc}$ and Eq.~\eqref{eq_pi4pi5}, we get 
    \begin{align}
        \left(\frac{\delta}{\lambda_\text{pc}}\right)^{3} \sim
    \frac{g}{\delta} \frac{w_\text{s}R_\text{s}^2}{U^3}\frac{\Delta\theta}{\theta_\text{a}}
    \implies
    \left(\frac{\delta}{\lambda_\text{pc}}\right)^{2} \sim
    \frac{gR_\text{p}}{U^2} \frac{\Delta\theta}{\theta_\text{a}}\frac{R_\text{p}}{\delta}  \implies \left(\frac{\delta}{\lambda_\text{pc}}\right) \sim \sqrt{\frac{\pi_3\pi_4}{\pi_2\pi_5}}.
    \end{align}

    
\noindent In this case as well, we have demonstrated the emergence of the $\pi$ groups that we have obtained in Eq.~\eqref{eq_pi_nocanopy} in the expression for the power-law functional dependency of the plume centerline slope.

\subsubsection{Unifying choice of $R_\text{p}$ across varied source geometries} \label{sect_hydradius}
For point, line, and finite (circular) sources, we have been able to show that the plume centerline slope in the rise phase can be expressed in terms of our $\pi$ groups. However, the cases differ in the power-law dependencies and combination of $\pi$ groups. Notably, the slope is independent of $\pi_2$ in the case of the line source, while it varies inversely with $\sqrt{\pi_2}$ and $\pi_2$ for point/finite sources when $\overline{z}<<L_\text{B}$ and $\overline{z}>>L_\text{B}$, respectively. This dependence of the slope on the non-dimensionalized ABL height ($\pi_2$) is attributed to the differences in the definition of the buoyancy-source strength across the two scenarios. We can propose a definition for the buoyancy-source strength that unifies the scaling for the point/finite and line sources as follows: 
\begin{align}
    B \sim g\frac{\Delta\theta}{\theta_\text{a}}w \times R_\text{p},~\text{where}~R_\text{p} = \frac{A_\text{s}}{P_\text{s}}. 
    \label{eq_hydradius}
\end{align}
Here, $A_\text{s}$ and $P_\text{s}$ are the buoyancy-source area and perimeter respectively. The dimensions of buoyancy source strength are $L^3T^{-3}$, consistent with those for a line source. If a line source is conceived as a rectangle (of sides $L_\text{x},~L_\text{y}$) with an extremely high aspect ratio ($L_\text{y}>>L_\text{x}$), 
\begin{align*}
    R_\text{p} = \frac{L_\text{x}L_\text{y}}{2(L_\text{x}+L_\text{y})} \approx\frac{L_\text{x}}{2}~\text{as}~\frac{L_\text{y}}{L_\text{x}}\rightarrow\infty.
\end{align*}
This makes Eq.~\eqref{eq_hydradius} consistent with Eq.~\eqref{eq_buoyancy_line}. For a point or finite circular source, $ R_\text{p} = \pi R_\text{s}^2/2\pi R_\text{s}= R_\text{s}/2$, where $R_\text{s}$ is defined as above. In this manner, a source length scale $R_\text{p}$ can be defined for any source geometries ranging from a point to a finite source to a line source. With this unified definition of buoyancy source strength, dimensional considerations, such as those employed by Raupach \cite{raupach1990similarity}, suggest that $\overline{z}/x \sim B^{1/3}/U$ for each of these geometries. Following the procedure outlined in Sect.~\ref{sect_line_nocan} to obtain the average plume centerline slope for all these scenarios, we get the same expression as in Eq.~\eqref{eq_slope_line}:
\begin{align}
    \frac{\delta}{\lambda_\text{pc}} \sim \sqrt{\frac{gR_\text{p}}{U^2}\frac{\Delta\theta}{\theta_\text{a}}} \equiv \sqrt{\frac{\pi_3\pi_4}{\pi_5}}.
    \label{eq_slope_unified}
\end{align}

\subsubsection{Infinite plane source, cross-wind, no canopy}
We can now attempt to obtain a scaling for the average plume centerline slope for an infinite-plane source of buoyancy in cross-wind, which was precluded in Raupach's analysis \cite{raupach1990similarity}. In this case, $R_\text{p}\rightarrow \infty$ and we define the relevant length scale to be $\lambda_\text{pc}$. Based on the explanation provided in Sect.~\ref{sect_hydradius}, the buoyancy-source strength is given by:
\begin{align*}
    B \sim g\frac{\Delta\theta}{\theta_\text{a}}w \lambda_\text{pc}.
    \end{align*}
  $\text{Using} ~\overline{z}/x \sim B^{1/3}/u \sim w/u \sim \delta/\lambda_\text{pc}$, {\text{we get}:
  \begin{align}
\left(\frac{\overline{z}}{x}\right)^3 \sim \left(\frac{\delta}{\lambda_\text{pc}}\right)^3\sim\frac{B}{u^3} \sim    \frac{g}{u^2}\frac{\Delta\theta}{\theta_\text{a}}\frac{w}{u} \lambda_\text{pc} \sim  \frac{g}{u^2}\frac{\Delta\theta}{\theta_\text{a}}\frac{\delta}{\lambda_\text{pc}} \lambda_\text{pc}
    \implies \frac{\delta}{\lambda_\text{pc}} \sim \left( \frac{g\delta}{U^2}\frac{\Delta\theta}{\theta_\text{a}}\right)^{1/3} \equiv \left[\pi_2\pi_3\frac{\pi_4}{\pi_5}\right]^{1/3}.
    \label{eq_infplane}
\end{align},
    
Again, the average plume centerline slope is obtained in terms of our $\pi_\text{i}$ groups. \textcolor{black}{Note that with an increase in the ABL height ($\delta$), $\pi_2\pi_3$ increases; this means that the average plume centerline slope increases (plume deflects less from the vertical) with a deeper ABL for a constant ambient velocity scale. 
This is explained by the fact that the ambient wind shear (scaling), given by $U/\delta$, is stronger (higher) for a shallower ABL; this leads to increased plume tilting from the vertical for a constant ambient velocity scaling, resulting in a more gradual plume centerline slope when the ABL is shallower.} Moreover, this is also consistent with deeper ABLs being associated with stronger convection.

\subsubsection{Canopy, cross-wind, finite source}
A unique study on buoyant-plume interaction with a vegetative canopy in cross-wind was conducted by Chung and Koseff \cite{chung2023interaction} using flume experiments. They obtained a power-law relationship between the coefficient of variation in plume centerline trajectory ($s_\mu/\overline{z}$) and a (modified) Byram's convective number, in which the velocity scaling was taken to be $u_*$ as opposed to $U$:
\begin{equation*}
\frac{s_\mu}{\overline{z}} \sim \left(\frac{B}{u_*^3}\right)^{\beta},~\beta<0.  
\end{equation*}
Here, $s_\mu$ is the standard deviation of the plume centerline computed at a location in the far-field region \cite{chung2023interaction}. In their analysis, $s_\mu/\overline{z}$ was also plotted as a function of the bulk Richardson number ($Ri_\text{b}$). They found that the trends were similar to those observed with $B/u_*^3$ and consistent with a power-law scaling. Therefore, 
\begin{align}
    \frac{s_\mu}{\overline{z}} \sim Ri_\text{b}^\alpha, \alpha < 0  \implies \frac{s_\mu}{\overline{z}} &\sim \left[g\beta_\text{T}\frac{\Delta\theta}{L_\text{pl}}\left(\frac{L_\text{RS}}{u_*}\right)^2\right]^\alpha \implies \left(\frac{B}{u_*^3}\right)^{\beta} \sim \left[g\beta_\text{T}\frac{\Delta\theta}{L_\text{pl}}\left(\frac{L_\text{RS}}{u_*}\right)^2\right]^\alpha. 
\label{eq_canopy_rib}
  \end{align}
Here, we are using the definition of $Ri_\text{b}$ by Chung and Koseff \cite{chung2023interaction}, $L_\text{pl}$ is a plume length scale, and $L_\text{RS}$ is the shear length scale. 
Established theory on canopy turbulence can be used to recast some of the physical quantities above in terms of the current variables under consideration: 

\begin{enumerate}[label = (\roman*)]
    \item \textbf{Plume length scale} ($L_\text{pl}$): 
According to Chung and Koseff \cite{chung2023interaction}, the frequency of Kelvin-Helmholtz (KH) rollers near the canopy top ($f_\text{KH}$) dictates plume centerline oscillation. They define $L_\text{pl}$ as the vertical length traversed by the plume within one KH period ($f_\text{KH}^{-1}$). The KH frequency is determined by the peak spectral frequency of the vertical velocity ($f_\text{p}(w)$), which is given by \cite{raupach1996coherent}:
\begin{align*}
  f_\text{KH} \simeq  f_\text{p}(w) = \frac{U_\text{c}}{\Lambda_\text{x}} \approx \frac{U}{C_0 h_\text{c}},
    \end{align*}
where $\Lambda_\text{x}$ represents the horizontal wavelength of KH rollers and $U_\text{c}$ represents their convection velocity. Using the relation $\Lambda_\text{x} = C_0 h_\text{c}$ is obtained from the mixing layer analogy used by Raupach \cite{raupach1996coherent}, 
\begin{align*}
    L_\text{pl} = w\times f_\text{KH}^{-1} = w\times \frac{C_0 h_\text{c}}{U}.
\end{align*}

\item \textbf{Shear length scale} ($L_\text{RS}$): The shear length scale, i.e. the height from the canopy top over which the turbulent stress remains substantial, has been found to be proportional to the canopy height, as indicated by data across different vegetation heights \cite{raupach1996coherent}. Hence, $L_\text{RS} = C_1 h_\text{c}$.
\end{enumerate}

\noindent Substituting for these terms in Eq.~\eqref{eq_canopy_rib}, dropping the constants $C_0,~ C_1$, and using Eq.~\eqref{eq_pi4pi5} and $w/U \sim \delta/\lambda_\text{pc}$, 
we get: 
\begin{align}
    &\left[\frac{B}{u_*^3}\right]^{\beta/\alpha} \sim g\frac{\Delta\theta}{\theta_\text{a}}\left(w\frac{C_0 h_\text{c}}{U}\right)^{-1}\left(\frac{L_\text{RS}}{u_*}\right)^2,~
    \text{where}~B \simeq g \frac{\Delta\theta}{\theta_\text{a}}wR_\text{p}, \notag\\
    &\implies \left[\frac{g}{u_*^2} \frac{\Delta\theta}{\theta_\text{a}}\frac{w}{u_*} R_\text{p}\right]^{\beta/\alpha} \sim ~\frac{g}{u_*^2}\frac{\Delta\theta}{\theta_\text{a}}\left(w \frac{h_\text{c}}{U}\right)^{-1}(h_\text{c})^2
    \implies \left[\frac{gR_\text{p}}{U^2} \frac{\Delta\theta}{\theta_\text{a}}\frac{w}{U}\frac{U^3}{u_*^3}\right]^{\beta/\alpha} \sim  ~\frac{gR_\text{p}}{U^2}\frac{U^2}{u_*^2}\frac{\Delta\theta}{\theta_\text{a}}\left(\frac{w}{U}\right)^{-1}\left(\frac{h_\text{c}}{R_\text{p}}\right)  \notag\\
    &\implies    \left(\frac{\delta}{\lambda_\text{pc}}\right)^{1+\frac{\beta}{\alpha}} \sim (\pi_3\pi_4)^{1-\frac{\beta}{\alpha}}\pi_5^{4\frac{\beta}{\alpha}-3}\pi_6 \implies \frac{\delta}{\lambda_\text{pc}} \sim 
    \left[(\pi_3\pi_4)^{\alpha-\beta}\pi_5^{4\beta-3\alpha}\pi_6^{\alpha}\right]^{\frac{1}{\alpha+\beta}}
    \label{eq_slope_canopy_chung}
\end{align}

From power-law fits to the plot of $s_\mu/\overline{z}_\text{pc}$ against the modified Byram's convective number for various buoyancy-source strength and wind speeds, Chung and Koseff obtained $\beta = -0.22$. From a power-law fit to the plot of $s_\mu/\overline{z}_\text{pc}$ against $Ri_\text{b}$ provided by Chung and Koseff, obtained using a simplex algorithm (MATLAB's \textit{fminsearch}() function), we get $\alpha = -0.19$ with a coefficient of determination, i.e. $R^2=0.53$ (process not shown here). Therefore, Eq.~\eqref{eq_slope_canopy_chung} becomes:

\begin{equation}
    \frac{\delta}{\lambda_\text{pc}} \sim  (\pi_3\pi_4)^{-0.07}\pi_5^{0.76}\pi_6^{0.46}
    \label{eq_slope_can_chung2}
\end{equation}
Equation~\eqref{eq_slope_can_chung2} depicts the dependence of the plume centerline slope on our $\pi$ groups as we intended. It is worth noting that the dependence on $\pi_3\pi_4$ is inverse and weaker compared to the dependence on $\pi_5$ and $\pi_6$, suggesting that the slope is a stronger function of the canopy geometry and turbulence intensity. Albeit weak, the inverse dependence is inconsistent with an expected increase in slope for an increase in $\Delta H_\text{s}/\rho c_\text{p}$. Modeling the canopy using dowel rods and releasing the plume at the canopy height also results in the preclusion of canopy density impacts, i.e of $\pi_7$, on the plume slope. Moreover, the exponents in Eq.~\eqref{eq_slope_can_chung2} were obtained from the coefficient of variation in the plume centerline in the far-field region, making it difficult to characterize the plume slope closer to the source. 
The scale of the experiment, i.e. the shallow depth of the water tank relative to the canopy height (4$h_\text{c}$), insufficient data points for the power-law fit, and a narrow range of wind speeds constitute other potential limitations of the analysis in the context of determining scaling laws. These limitations emphasize the need for additional experiments and computer simulations to characterize plume deflection both near the source and in the far-field regions. 

\subsection{Plume centerline slope and transition height: scaling from LES}
\label{sect_les_scaling}
In this section, we investigate the dependence of the 1-h mean plume centerline slope ($\Delta\overline{z}/\Delta x$) on the $\pi$ groups in Eqs.~\eqref{eq_pi_canopy} and \eqref{eq_pi_nocanopy}. Figure~\ref{fig_slope_zones} presents the 1-h mean vertical velocity in the streamwise ($XZ$) plane passing through the spanwise center of the square surface heat source, for $H_\text{patch}=5$\,kW\,m\textsuperscript{-2} and $U=2$\,m\,s\textsuperscript{-1}, with $h_\text{c}=35$\,m and $c_\text{d}LAI=1$ in the canopy case. The plume centerline slope is tracked from the locus of the maximum 1-h mean vertical velocity ($\overline{w}_\text{max}$) on this plane.  In the region near the plume source (Zone 1), buoyancy dominates the ambient wind forcing and $\overline{w}_\text{max}$ is obtained from the maximum $\overline{w}$ across a horizontal transect in this plane at different altitudes. In Zone 1, $\Delta\overline{z}/\Delta x$ is steeper, indicating that the plume centerline is close to vertical, albeit to varying degrees, across the parametric space considered in this study. In the far field (Zone 2), the plume centerline flattens and the influence of the ambient cross-wind strengthens relative to buoyancy effects. Here, the centerline $\overline{w}_\text{max}$ is obtained from the maximum $\overline{w}$ across a vertical transect in this plane at different streamwise locations. The transition between the two zones is determined by tracking a sudden change in the local plume centerline slope beyond a certain threshold, in Zone 1, as demonstrated in Fig.~\ref{fig_slope_zones}(a). Note that the two-zone plume centerline is particular to the no-canopy case. In the presence of a canopy, there exists an additional intermediate zone (Zone 1(B)), i.e. the region where the plume, informed by its interaction with the canopy, adjusts to the ambient cross-wind forcing (Fig.~\ref{fig_slope_zones}(b)). More information about the plume centerline behavior in these regions is provided by Desai et al. \cite{desai2025investigating}.


\begin{figure}[h!]
    \centering
    \includegraphics[scale=0.35]{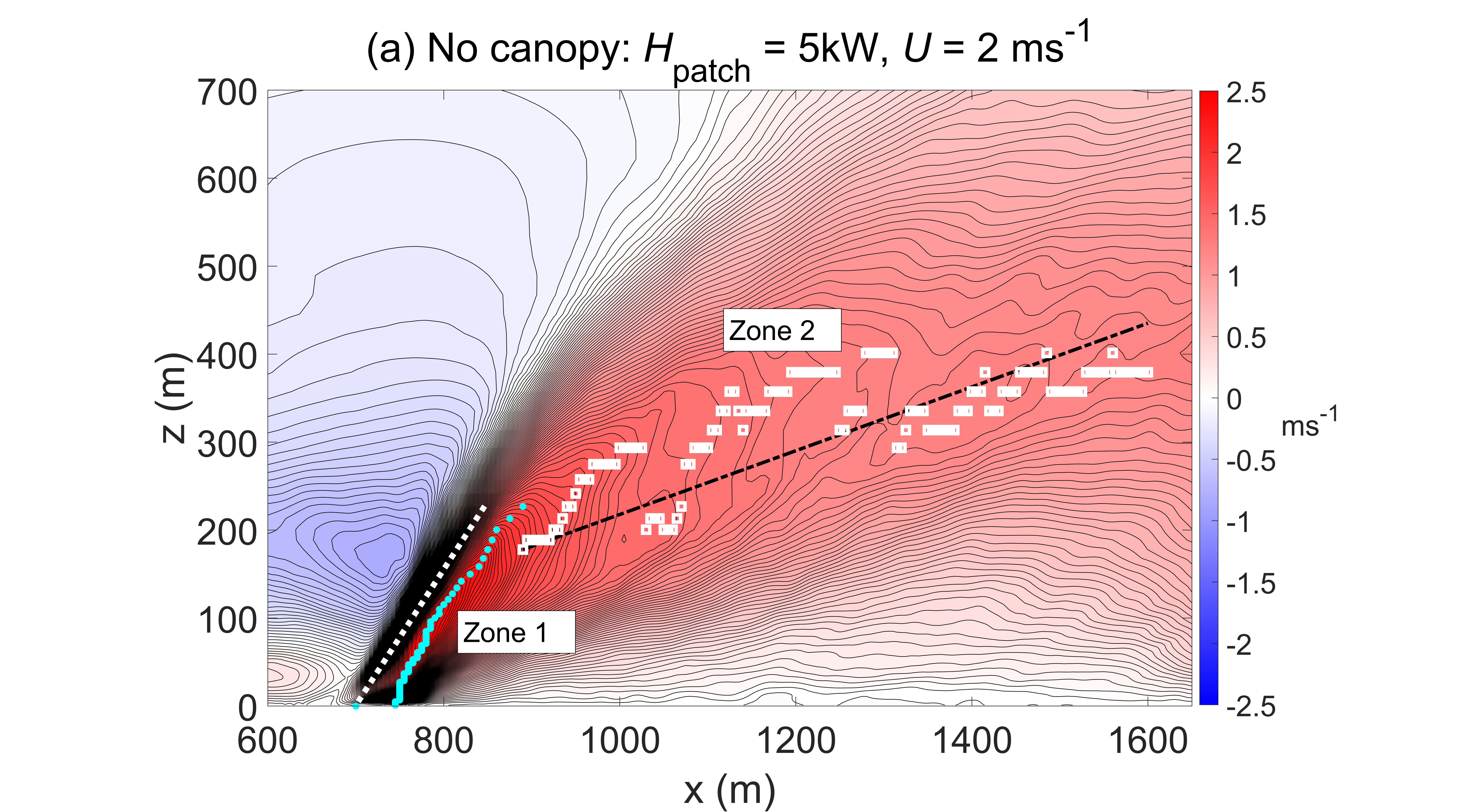}\\
    \includegraphics[scale=0.35]{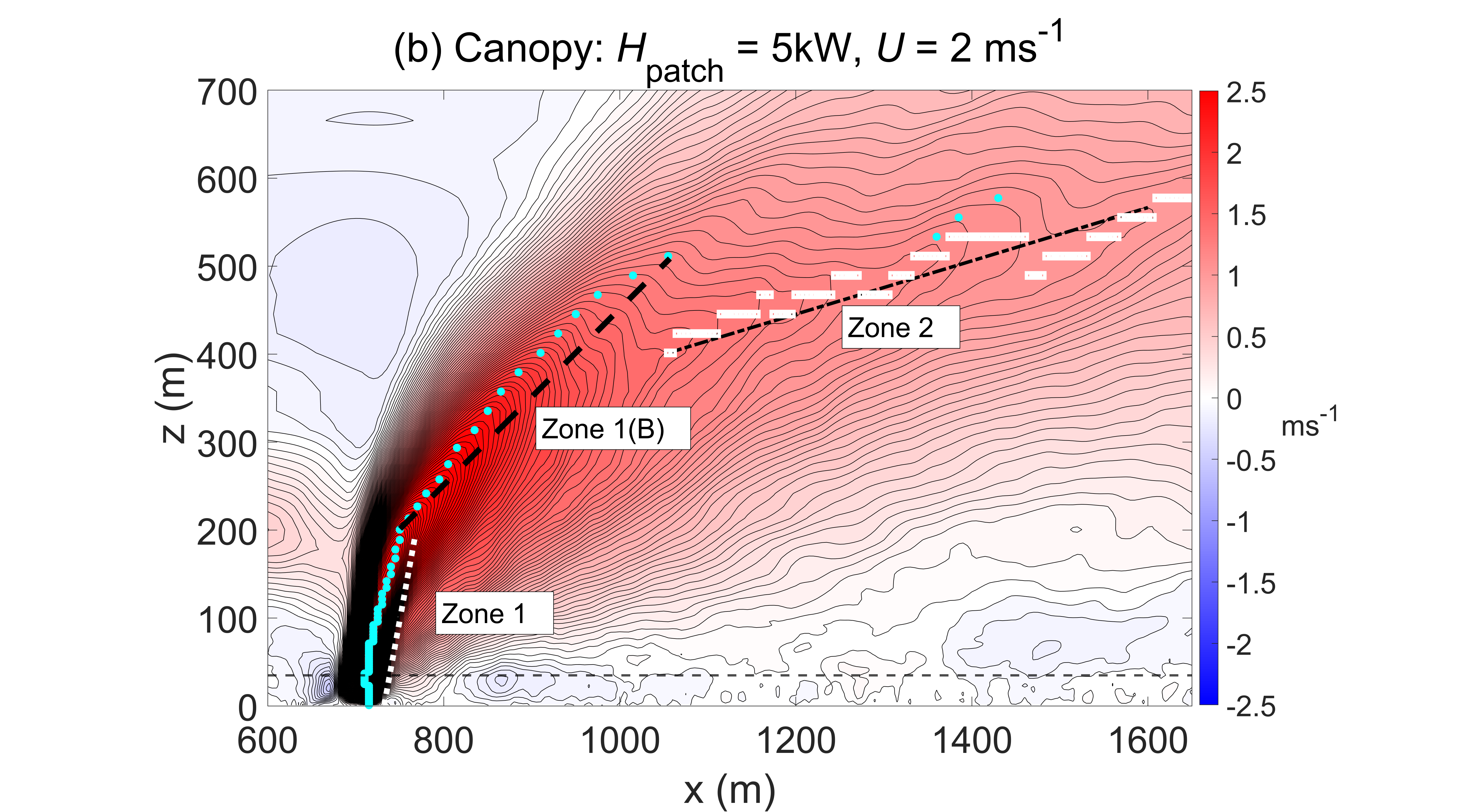}
    \caption{Color contours of the 1-h mean vertical velocity ($\overline{w}$) for the (a) no-canopy and (b) canopy environments depicting the respective plume centerline inclinations in each zone. Trend-lines in Zone 2 are obtained using MATLAB's \textit{fminsearch}() for a linear fit to the oscillating mean plume centerline as done by Desai et al. \cite{desai2025investigating}. Blue dots and white squares represent the plume centerline in Zones 1-1(B) and Zone 2, respectively}
    \label{fig_slope_zones}
\end{figure}

A power-law dependence is explored between the plume centerline slope and the derived dimensionless parameters as follows: 
\begin{equation}
    \frac{\Delta\overline{z}}{\Delta x} \sim \left(\frac{gR_\text{p}}{U^2} \frac{\Delta H_\text{s}}{\rho c_\text{p}}\frac{1}{U\theta_\text{a}}\right)^{\hat{p}}\left(\frac{u_*}{U}\right)^{\hat{r}} \left(\frac{h_\text{c}}{R_\text{p}}\right)^{\hat{s}_1}\left(\frac{L_\text{c}}{R_\text{p}}\right)^{\hat{s}_2}.
    \label{eq_pc_can1}
\end{equation}
Here, the first term in parenthesis on the right-hand side is $\pi_3\pi_4$, where $\pi_3$ and $\pi_4$ are grouped together based on the recurring occurrence of such grouping under the Clark
convective Froude number (or equivalently, Byram's convective number) in the literature and as shown in Sect.~\ref{sect_liter_scaling} above. When considering the next $\pi$ term, i.e. $\pi_5 = u_*/U$, the choice of location for $u_*$ in the flow field matters. Since $u_*$ ($\sqrt{|\overline{u'w'}|}$) varies spatially, i.e. upstream and downstream of the plume, it is important to consider contributions from both sides. Figure~\ref{fig_ustarloc} represents the momentum fluxes computed on the streamwise ($XZ$) plane passing through the spanwise center of the heat source for a no-canopy case and canopy case in similar forcing conditions ($H_\text{patch}=5$\,kW\,m\textsuperscript{-2}, $U=2$\,m\,s\textsuperscript{-1}, $h_\text{c}=35$\,m). For this case, Desai et al. \cite{desai2025investigating} discussed the influence of the spatial structure of the momentum fluxes on the plume centerline slope close to the surface. The positive momentum fluxes within the plume, on the leeward side of the plume centerline (black dots in Fig.~\ref{fig_ustarloc}(b)) in the canopy case, represent the upflux of momentum from the buoyancy source to the atmosphere aloft, and serve to keep the plume more upright. In contrast, the plume tilt is higher in the no-canopy case, where momentum fluxes are negative on the leeward side of the plume centerline (black dots in Fig~\ref{fig_ustarloc}(a)) and represent the downflux of momentum from the atmosphere aloft. It is, therefore, expected that the intensity of these leeward momentum fluxes relative to those from an upstream location will have some bearing on the plume centerline tilt for all the simulations in the current study. 
This motivates the addition of a supplementary $\pi$ term, $\hat{\pi}_{5}$, such that $\hat{\pi}_{5} =u_*^{(\text{up})}/U$, where $u_*^{(\text{up})}$ represents the friction velocity at an upstream location.

For the friction velocity in $\pi_5$ ($u_*^\text{(pl)}$), we first determine the leeward edge of the plume, i.e. where the mean vertical velocity approaches approximately $\overline{w}_\text{max}e^{-1}$ (magenta and cyan squares in Figs.~\ref{fig_ustarloc}(a) and \ref{fig_ustarloc}(b), respectively), where $\overline{w}_\text{max}$ represents the mean vertical velocity at the plume centerline as discussed above. We then choose the first grid point above the canopy top ($z=37$\,m,) on this edge for $u_*^\text{(pl)}$. Similarly, $u_*^{(\text{up})}$ also corresponds to the first grid point above the canopy, i.e. $z=37$\,m, and is taken at a streamwise location 325\,m upstream ($x=375$\,m) from the streamwise center of the heated surface patch ($x=700$\,m). This location was selected to provide sufficient fetch for the inlet wind profile to adjust to the presence of the canopy. We consider $x=375$\,m to be appropriate since it is about three times longer than the longest canopy adjustment length scale simulated ($L_\text{c}=127.5$\,m, Sect~\ref{sect_methods}).  Equation~\ref{eq_pc_can1} is now rewritten as: 
\begin{align}
\begin{split}
     \frac{\Delta\overline{z}}{\Delta x} \sim& \left(\frac{gR_\text{p}}{U^2} \frac{\Delta H_\text{s}}{\rho c_\text{p}}\frac{1}{U\theta_\text{a}}\right)^{\hat{p}}\left(\frac{\pi_5}{\hat{\pi}_{5}}\right)^{\hat{r}} \left(\frac{h_\text{c}}{R_\text{p}}\right)^{\hat{s}_1}\left(\frac{L_\text{c}}{R_\text{p}}\right)^{\hat{s}_2}
        \sim \left(\frac{gR_\text{p}}{U^2} \frac{\Delta H_\text{s}}{\rho c_\text{p}}\frac{1}{U\theta_\text{a}}\right)^{\hat{p}}\left(\frac{u_*^{(\text{pl})}/U}{u_*^{(\text{up})}/U}\right)^{\hat{r}} \left(\frac{h_\text{c}}{R_\text{p}}\right)^{\hat{s}_1}\left(\frac{L_\text{c}}{R_\text{p}}\right)^{\hat{s}_2}\\ &\equiv \left(\frac{gR_\text{p}}{U^2} \frac{\Delta H_\text{s}}{\rho c_\text{p}}\frac{1}{U\theta_\text{a}}\right)^{\hat{p}}\left(\frac{u_*^{(\text{pl})}}{u_*^{(\text{up})}}\right)^{\hat{r}} \left(\frac{h_\text{c}}{R_\text{p}}\right)^{\hat{s}_1}\left(\frac{L_\text{c}}{R_\text{p}}\right)^{\hat{s}_2}.
\end{split}
    \label{eq_pc_can2}
\end{align}

\begin{figure}
    \centering
    \begin{tabular}{cc}
    (a) No canopy & (b) Canopy\\
    \includegraphics[width=0.42\linewidth, trim={0 0 0 5mm}, clip]{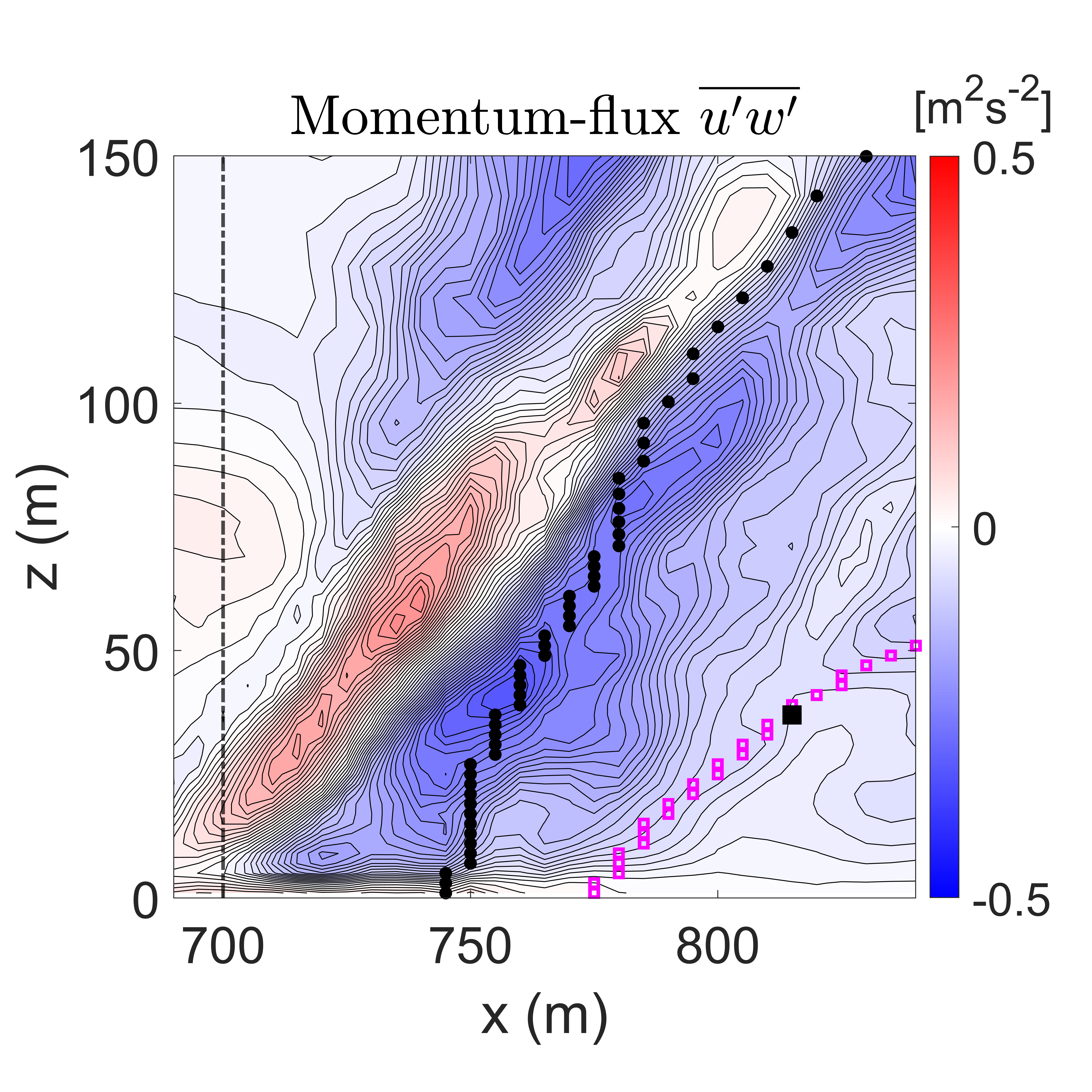}     &  \includegraphics[width=0.42\linewidth, trim={0 0 0 5mm}, clip]{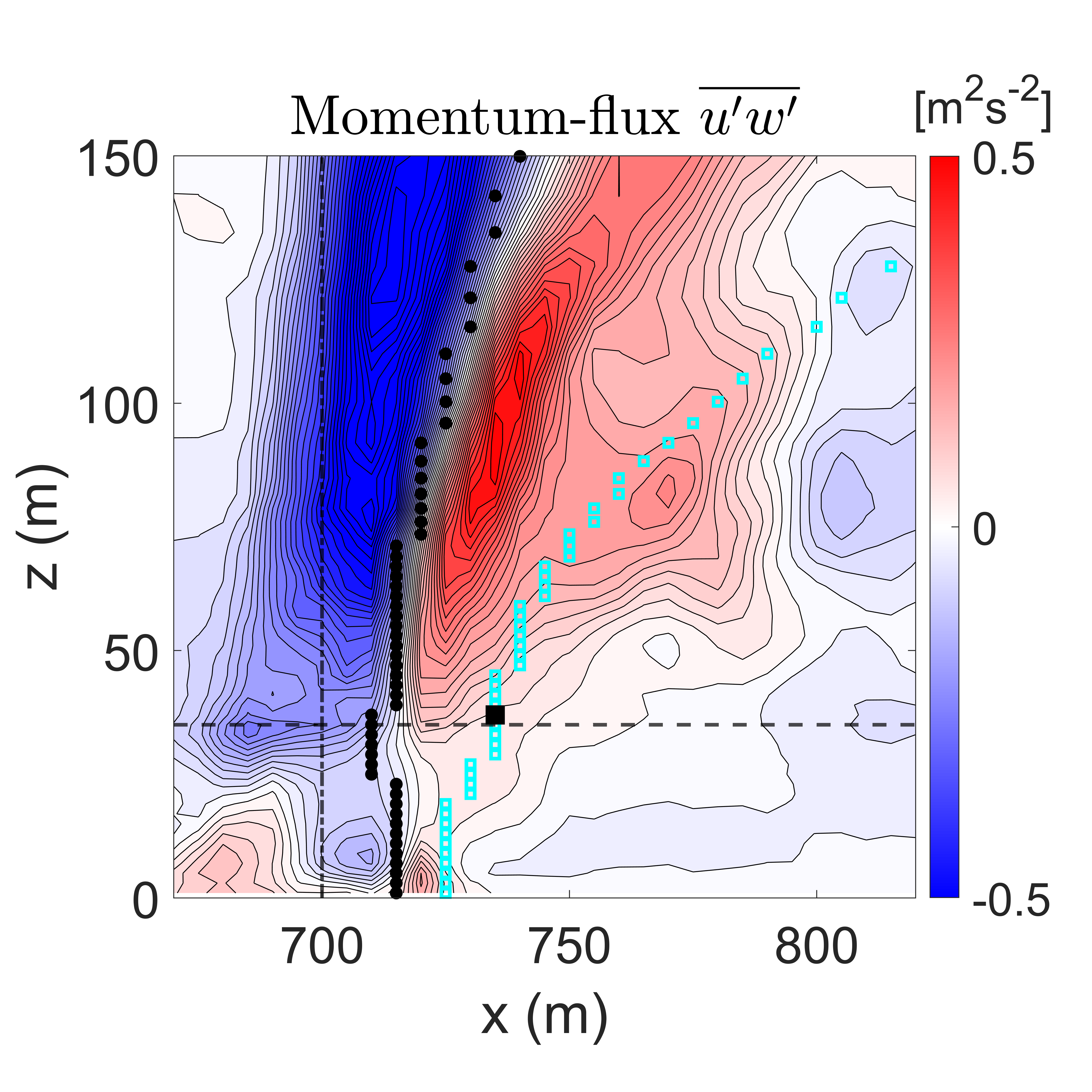}
    \end{tabular}
    \caption{A comparison of the near-source momentum fluxes, plotted on the streamwise ($XZ$) plane passing through the spanwise center of the heat source, between the (a) no-canopy and (b) canopy settings for the base case discussed in Sect.~\ref{sect_methods}. Magenta and cyan squares represent the leeward plume edge in (a) and (b), respectively. Black dots and the black square represent the plume centerline and the selected location for $u_*$ in both (a) and (b)}
    \label{fig_ustarloc}
\end{figure}

Moreover, $\hat{s}_1,~\hat{s}_2 \rightarrow 0$, in the no-canopy cases. We first explore the plume centerline dependence on $\pi_3\pi_4$ and $\pi_5$ in both environments, in Sect.~\ref{sect_pi3pi4pi5}, and then the dependence on $\pi_5$ and $\pi_6$ in Sect.~\ref{sect_pi5pi6} for the canopy case.

\subsubsection{Canopy and No-Canopy: Dependence on buoyancy source strength versus ambient forcing, plume turbulence intensity}
\label{sect_pi3pi4pi5}

\textbf{Plume centerline slopes.} Figure~\ref{fig_can_slopes_pi345}(a) shows a plot of $\Delta\overline{z}/\Delta x$ against $\pi_3\pi_4$ for Zones 1--2, with both axes scaled logarithmically. The power-law dependence between the two is represented in the form of a trend-line for each zone and takes the form: $ \left(\frac{\Delta\overline{z}}{\Delta x}\right)_\text{i} = \hat{m}_\text{i}\left(\frac{gR_\text{p}}{U^2} \frac{\Delta H_\text{s}}{\rho c_\text{p}}\frac{1}{U\theta_\text{a}}\right)^{\hat{p}_\text{i}}$, where the index i represents the zone number. It is observed that $\hat{m}_\text{1} > \hat{m}_\text{1(B)}$ by $\mathcal{O}(10^1)$ and $\hat{m}_\text{1} > \hat{m}_\text{2}$ by $\mathcal{O}(10^2)$, suggestive of the consistently steep plume centerline slopes near the surface in the canopy presence across the range of $\pi_3\pi_4$. Moreover, $\hat{p}_\text{1} \approx \hat{p}_\text{2} >> \hat{p}_\text{1(B)}$, suggesting a stronger dependency of the plume centerline slope on $\pi_3\pi_4$ near the canopy and plume source, in addition to the far-field region, as compared to the intermediate zone. In Zones 1 and 2, the plume centerline shows an an increasing trend with $\pi_3\pi_4$, with $\hat{p}_\text{1} \approx \hat{p}_\text{2} \approx 1/6$. The increasing trend is expected since higher values of $\pi_3\pi_4$ represent conditions in which buoyancy effects are stronger relative to the ambient forcing. 
In contrast, the plume centerline in the no-canopy cases follows a two-zone trajectory (Fig.~\ref{fig_slope_zones}(a)). Figure~\ref{fig_can_slopes_pi345}(b) shows a plot of $\Delta\overline{z}/\Delta x$ against $\pi_3\pi_4$ for Zones 1--2, with both axes scaled logarithmically. 
As expected, the centerline slopes are higher (steeper) close to the plume source, which is verified by the fact that $\hat{m}_\text{1} > \hat{m}_\text{2}$ by $\mathcal{O}(10^1)$. Increasing trends are observed for the slopes as a function of $\pi_3\pi_4$ in both zones, with similar power-law indices. 
Note that the power laws do not match those presented in Sect.~\ref{sect_liter_scaling}, where the plume centerline slope is shown to vary as a function of $\sqrt{\pi_3\pi_4}$.

The slopes in Zone 1, i.e. near the surface, are higher in the canopy versus the no-canopy cases, due to the higher drag afforded by the canopy to the rising plume. This is also reflected in the higher value of the coefficient in the canopy cases (compare Fig.~\ref{fig_can_slopes_pi345}(a) with Fig.~\ref{fig_can_slopes_pi345}(b)). However, the power-law indices for the trend-lines near the surface (Zone 1) are quite similar between the canopy and no-canopy cases. 
The far-field (Zone 2) slopes in the no-canopy cases follow a different power-law dependence than in the canopy cases, suggesting that the far-field plume in the canopy case follows a canopy-adjusted trajectory despite the distance from the canopy. 
This suggests that canopy presence at the surface has some \say{far-reaching} effects on the plume trajectory, which cannot be approximated via no-canopy boundary conditions.  Additionally, Fig.~\ref{fig_can_slopes_pi345}(a) depicts the plume centerline slopes in Zones 1, 1(B), and 2, for the heterogeneous canopy cases simulated by Cervantes et al. \cite{cervantes2026impact} (Sect.~\ref{sect_setup}). These data points are excluded from the regression and are found to deviate considerably from the current power-law model. However, the associated slopes are observed to increase with $\pi_3\pi_4$ for most of these outlier cases, as expected from physical considerations.\\


\begin{figure}[h!]
    \centering
    \begin{tabular}{cc}
 \hspace{-5pt}      \includegraphics[scale=0.35]{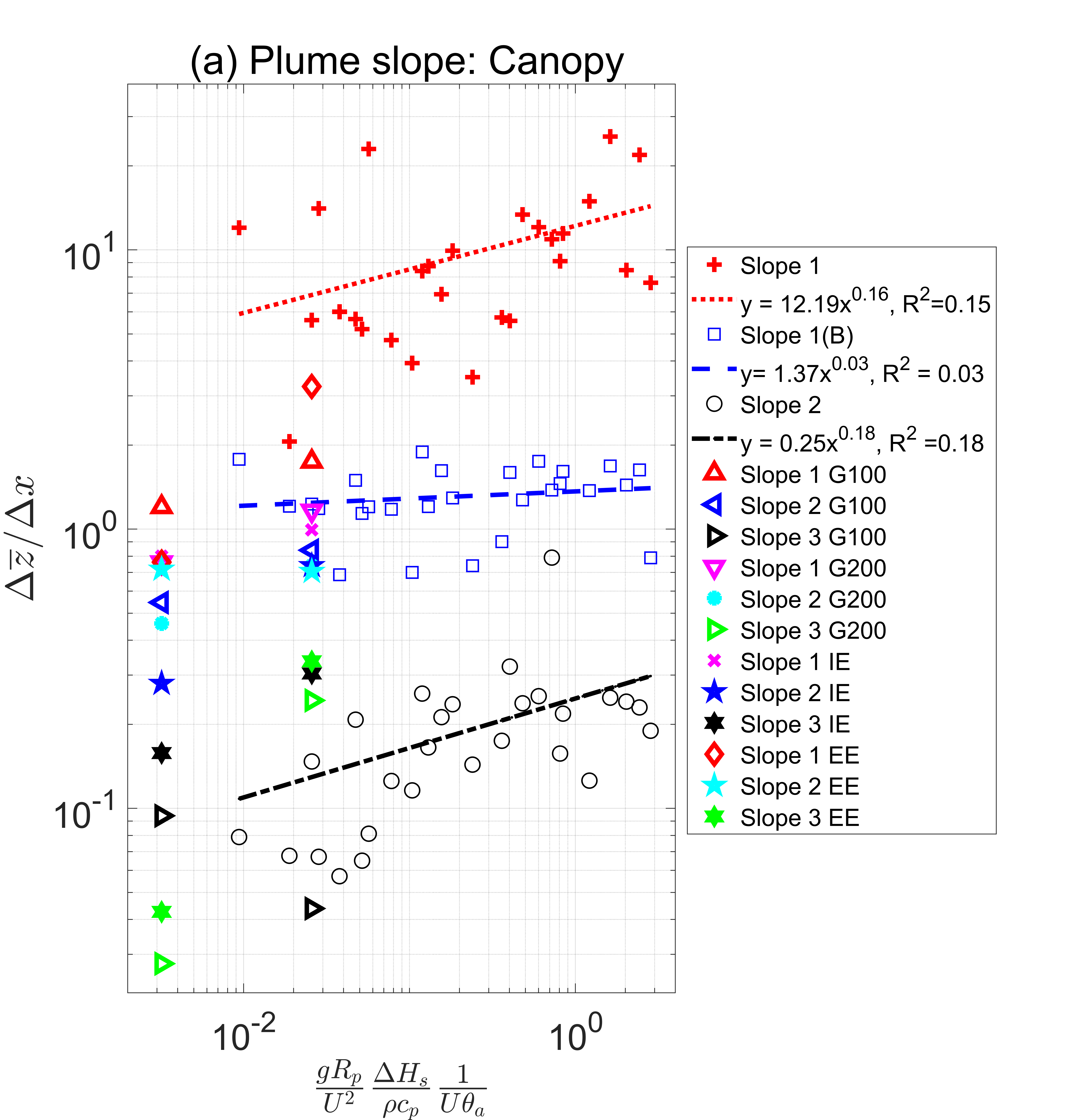}
        & \hspace{-20pt}    \includegraphics[scale=0.35]{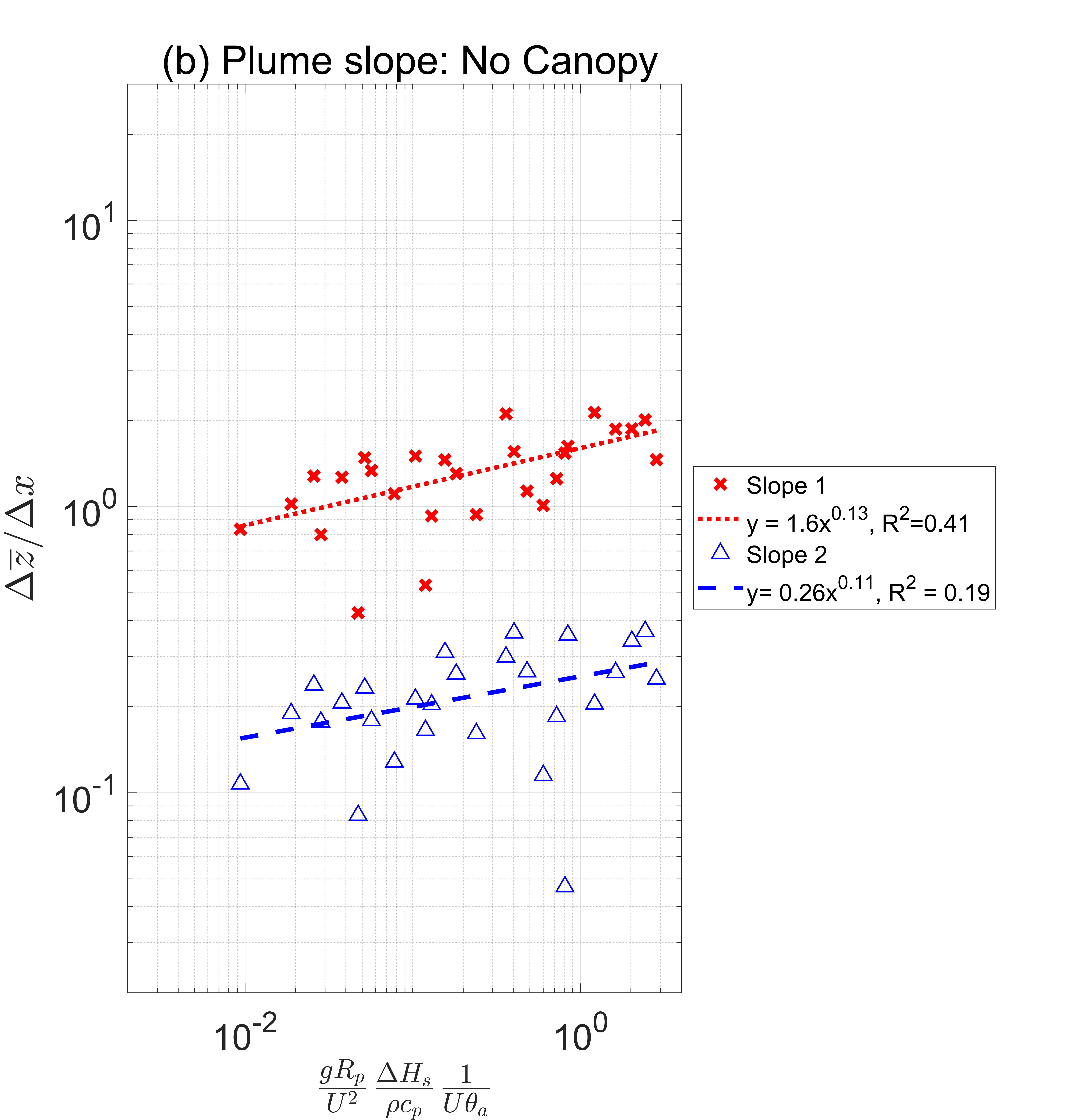}\\
         \hspace{-5pt}      \includegraphics[scale=0.35]{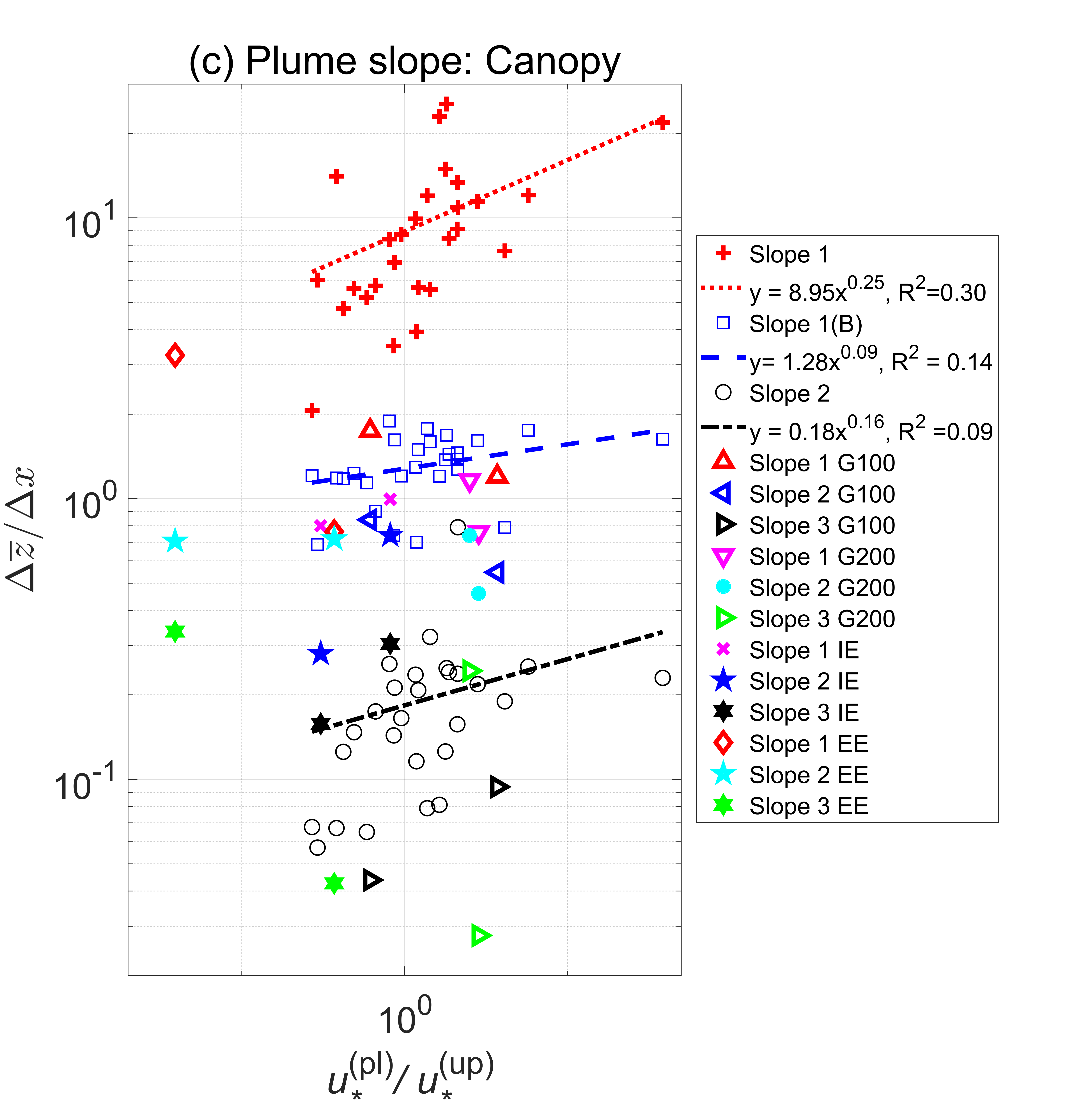}
        & \hspace{-20pt}    \includegraphics[scale=0.35]{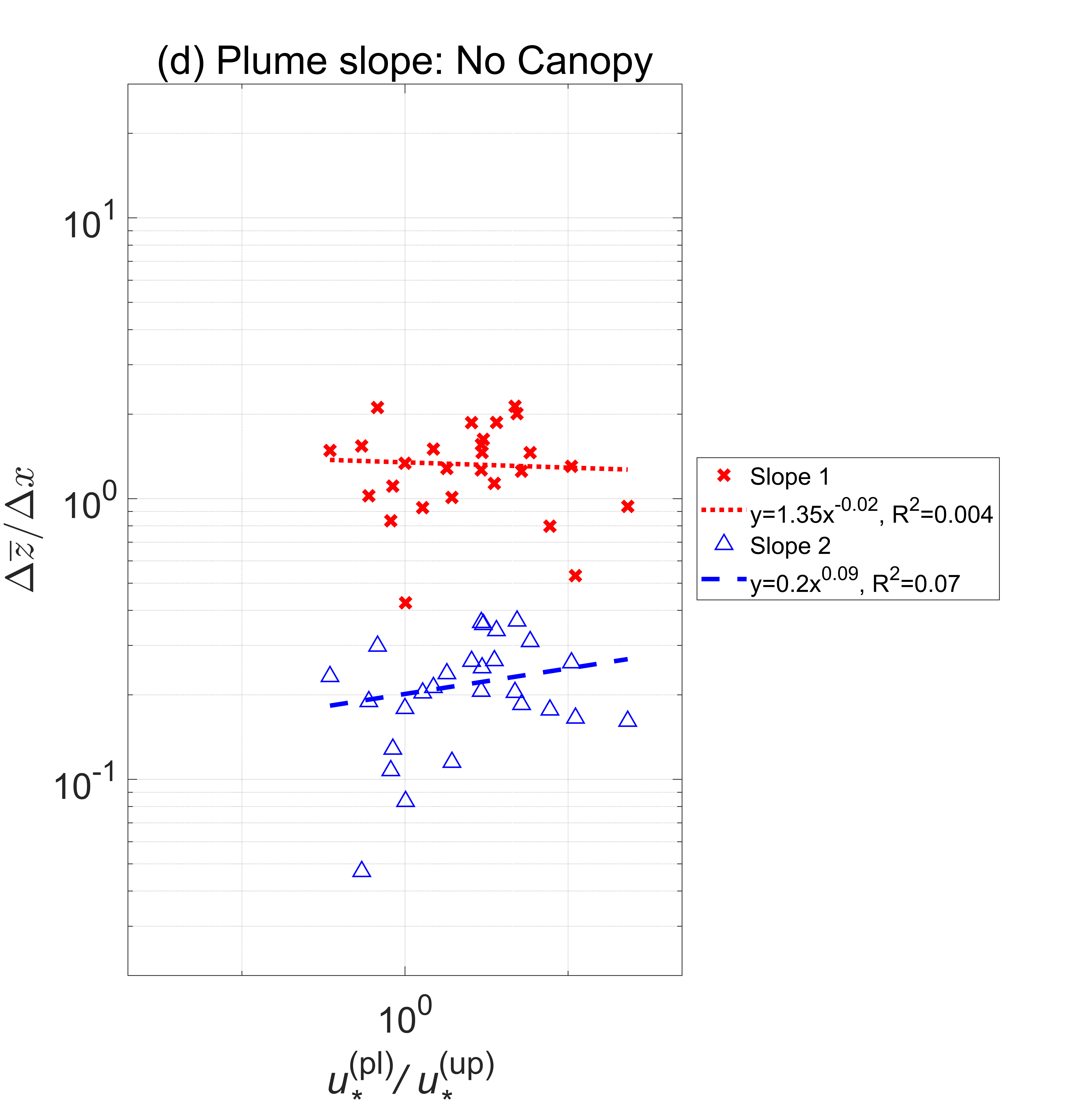}
    \end{tabular}
    \caption{Plot of plume centerline slope ($\Delta\overline{z}/\Delta x$) against $\pi_3\pi_4$ in the (a) canopy and (b) no-canopy cases, and against $u_*^{(\text{pl})}/u_*^{(\text{up})}$ in the (c) canopy and (d) no-canopy cases. 
    Slope i, computed from the simulation outputs, refers to the plume centerline slope in Zone i. Trend-lines represent the best fit of a power law to the data; $R^2$ represents the coefficient of determination}
    \label{fig_can_slopes_pi345}
\end{figure}

Figure~\ref{fig_can_slopes_pi345}(c) shows a plot of $\Delta\overline{z}/\Delta x$ against $u_*^{(\text{pl})}/u_*^{(\text{up})}$ for the canopy cases, with a trend-line that follows a power-law dependence.
Again, $\Delta\overline{z}/\Delta x$ changes more gradually with $u_*^{(\text{pl})}/u_*^{(\text{up})}$ in Zone 1(B), i.e. the intermediate zone, as seen from the low the power-law exponent. Trend-lines in Zones 1 and 2 show increasing trends in the plume centerline slope. The $1/4^\text{th}$ power law in Zone 1, which encompasses the canopy sublayer, demonstrates a stronger dependence on $u_*^{(\text{pl})}/u_*^{(\text{up})}$. This can be explained as follows. Higher values of $u_*^{(\text{pl})}/u_*^{(\text{up})}$ suggest a higher intensity of momentum transport from the buoyancy source within the canopy to the atmosphere aloft (positive momentum fluxes), relative to an upstream location where the plume influence is minimal and canopy-induced momentum-flux events, comprising mainly sweeps and ejections (negative momentum fluxes) that transport higher momentum downward, are more active. As explained earlier, a stronger upflux of momentum from the source on the leeward side contributes to a steeper plume centerline slope in Zone 1.
Figure~\ref{fig_can_slopes_pi345}(d) shows plots of $\Delta\overline{z}/\Delta x$ against $u_*^{(\text{pl})}/u_*^{(\text{up})}$ for the no-canopy cases in Zones 1--2, with trend-lines that each follow a power-law dependence. Interestingly, a decreasing trend is observed in the plume centerline slope with $u_*^{(\text{pl})}/u_*^{(\text{up})}$, in stark contrast to the canopy case (compare Fig.~\ref{fig_can_slopes_pi345}(c) with Fig.~\ref{fig_can_slopes_pi345}(d)). This, again, is a consequence of the difference in the spatial structure of the momentum fluxes within the plume in the canopy vs. no-canopy cases \cite{desai2025investigating} (Fig.~\ref{fig_ustarloc}). 
In the no-canopy case, higher values of $u_*^{(\text{pl})}/u_*^{(\text{up})}$ near the surface suggest more intense downward momentum transport, affording higher \say{drag} on the leeward side of the plume. We expect this to be associated with higher plume centerline deflection from the vertical and lower plume centerline slopes near the surface. Additionally, $\Delta\overline{z}/\Delta x$ has a higher power-law sensitivity to $u_*^{(\text{pl})}/u_*^{(\text{up})}$ in the canopy case compared to the no-canopy case, both near the surface (Zone 1) and in the far-field region (Zone 2). This suggests a stronger overall influence of canopy-influenced turbulence on the plume centerline behavior within the parametric space considered here. Furthermore, Fig.~\ref{fig_can_slopes_pi345}(c) depicts the plume centerline slopes in Zones 1, 1(B), and 2, for the heterogeneous canopy cases simulated by Cervantes et al. \cite{cervantes2026impact}. These data points are excluded from the regression. In addition, the dependence of $\Delta\overline{z}/\Delta x$  on $u_*^{(\text{pl})}/u_*^{(\text{up})}$ in these cases is more complex than the homogeneous canopy cases: the canopy structure informs both the upstream turbulence ($u_*^{(\text{up})}$) and momentum fluxes at the leeward edge of the plume ($u_*^{(\text{pl})}$), resulting in considerable deviations of the data points from the current power-law model. \\

\textbf{Plume transition altitude.} In addition to the plots on the plume centerline slopes above, we plot the height at which the plume transitions from the rise phase to the far-field bent-over phase, normalized by the ABL height ($z_\text{pc}/\delta$), in the canopy case (Zone 1(B) to Zone 2) and no-canopy case (Zone 1 to 2), as shown in Fig.~\ref{fig_ztrans_nocan}(a). It is observed that this height increases with an increase in $\pi_3\pi_4$ as evidenced both by the data from the LES and the trend-line indicating a $1/3^\text{rd}$ and $1/4^\text{th}$ power-law dependence in the no-canopy and canopy cases, respectively. This is expected, since an increase in buoyancy strength relative to the ambient wind forcing, i.e. an increase in $\pi_3\pi_4$, would result in a taller rise phase and delay the onset of the far-field bent-over phase. More importantly, the transition height in the canopy cases is consistently higher compared to the no-canopy cases, suggesting that the aerodynamic effect of the canopy delays the onset of the bent-over phase. This effect is more pronounced at lower values of $\pi_3\pi_4$, i.e. at weaker buoyancy source strengths. Canopy effects become less prominent at higher buoyancy source strengths, i.e. for $\pi_3\pi_4>1$, in which regime, plume transition altitudes are comparable with no-canopy conditions. Additionally, Fig.~\ref{fig_ztrans_nocan}(b) depicts $z_\text{pc}/\delta$ for the heterogeneous canopy cases simulated by Cervantes et al. \cite{cervantes2026impact}. These cases are excluded from the regression; they are observed to deviate from the power-law model for the current data to varying extents and exist as outliers. However, the normalized transition height for these is found to increase with $\pi_3\pi_4$, which is physically consistent.

We also plot the normalized buoyancy length scale ($L_\text{B}/\delta$) as a function of $\pi_3\pi_4$ in Fig.~\ref{fig_ztrans_nocan}(a), using the definition of $L_\text{B}$ from Eq.~\eqref{eq_traj_point_far}, which can be recast as: 
\begin{equation}
    L_\text{B} = g\frac{w_\text{s}R_\text{s}^2}{U^3}\frac{\Delta\theta}{\theta_\text{a}} \cong R_\text{p} \left(\frac{gR_\text{p}}{U^2} \frac{\Delta H_\text{s}}{\rho c_\text{p}}\frac{1}{U\theta_\text{a}}\right) \equiv R_\text{p}\pi_3\pi_4.  
    \label{eq_Lbpi3pi4}
\end{equation}
Here, we have used Eq.~\eqref{eq_pi4pi5} and $w_\text{s}\sim u_{*}$. Equation~\eqref{eq_Lbpi3pi4} shows that $L_\text{B}/\delta$ increases as a function of $\pi_3\pi_4$, which is corroborated by Fig.~\ref{fig_ztrans_nocan}. Furthermore, Fig.~\ref{fig_ztrans_nocan} shows that $L_\text{B}/\delta$ increases at a much faster rate compared to $z_\text{pc}/\delta$, so that the ratio between the transition height and the buoyancy length scale ($z_\text{pc}/L_\text{B}$) decreases with the increase in buoyancy strength. This is consistent with the scaling laws from previous studies discussed in Sect.~\ref{sect_liter_scaling}. For instance, Eqs.~\eqref{eq_nocan_far_transht} and \eqref{eq_nocan_near_transht} together suggest: 

\begin{align}
    \begin{split}
\left(\frac{\overline{z}}{x}\right)^3 \sim \frac{L_\text{B}}{\overline{z}} \text{ when } \overline{z}<<L_\text{B} \implies \frac{z_\text{pc}}{L_\text{B}} \sim \left(\frac{\delta}{\lambda_\text{pc}}\right)^{-3}, \\
\left(\frac{\overline{z}}{x}\right)^2 \sim \frac{L_\text{B}}{\overline{z}} \text{ when } \overline{z}>>L_\text{B} \implies \frac{z_\text{pc}}{L_\text{B}} \sim \left(\frac{\delta}{\lambda_\text{pc}}\right)^{-2}.
    \end{split}
    \label{eq_zpc_LB_scaling}
\end{align}

\noindent Here, we have used $z_\text{pc}/x_\text{pc} \equiv \delta/\lambda_\text{pc}$, which represents the plume centerline slope (Fig.~\ref{fig_schematic}). Equation~\eqref{eq_zpc_LB_scaling} suggests that while both $z_\text{pc}$ and $L_\text{B}$ increase with the buoyancy strength relative to ambient wind forcing ($\pi_3\pi_4$), the rate at which $z_\text{pc}$ increases is shallower compared to the rate at which $L_\text{B}$ increases, given the increase in the plume centerline slope ($\delta/\lambda_\text{pc}$) with $\pi_3\pi_4$. In other words, as the buoyancy strength increases relative to ambient wind forcing, the ratio of $z_\text{pc}$ to $L_\text{B}$ decreases. This is also an important observation in light of the differing power-law indices (Sect.~\ref{sect_liter_scaling}) based on whether $\overline{z}<<L_\text{B}$ or $\overline{z}>>L_\text{B}$. The assumption of $\overline{z}<<L_\text{B}$ in the rise phase of the plume becomes more reasonable when deriving the power-law scaling for the plume trajectory in this phase, at higher values of $\pi_3\pi_4$, while it becomes questionable at lower values of $\pi_3\pi_4$. This becomes an added consideration when determining the scaling law for the trajectory near the source when the buoyancy is weaker compared to the ambient wind forcing. This emphasizes the need for more studies to investigate the power-law dependencies across multiple scales.

\begin{figure}[h!]
    \centering
    \begin{tabular}{cc}
     \includegraphics[scale=0.45, trim={0 0 18mm 0}, clip]{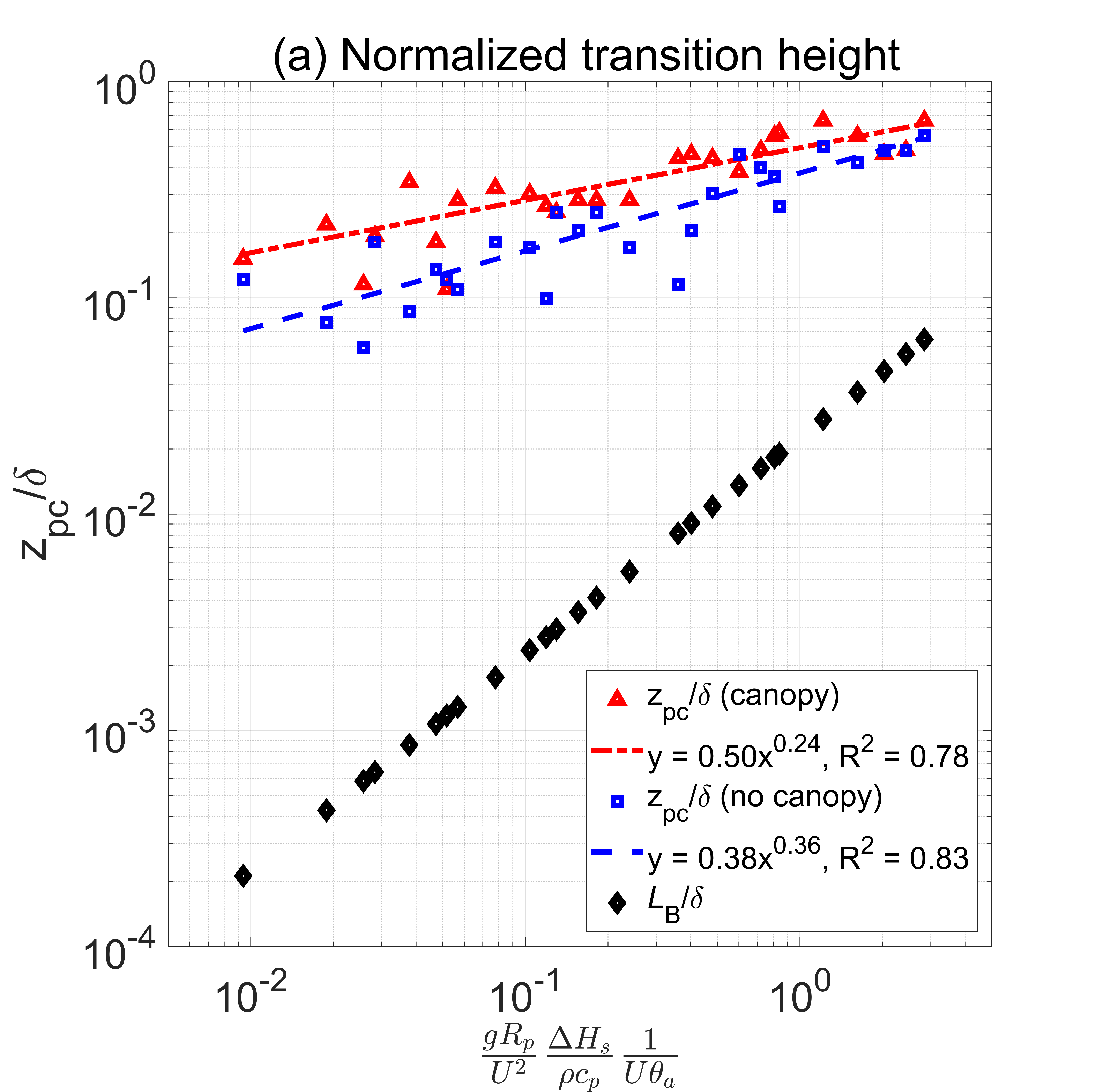}    & \includegraphics[scale=0.45]{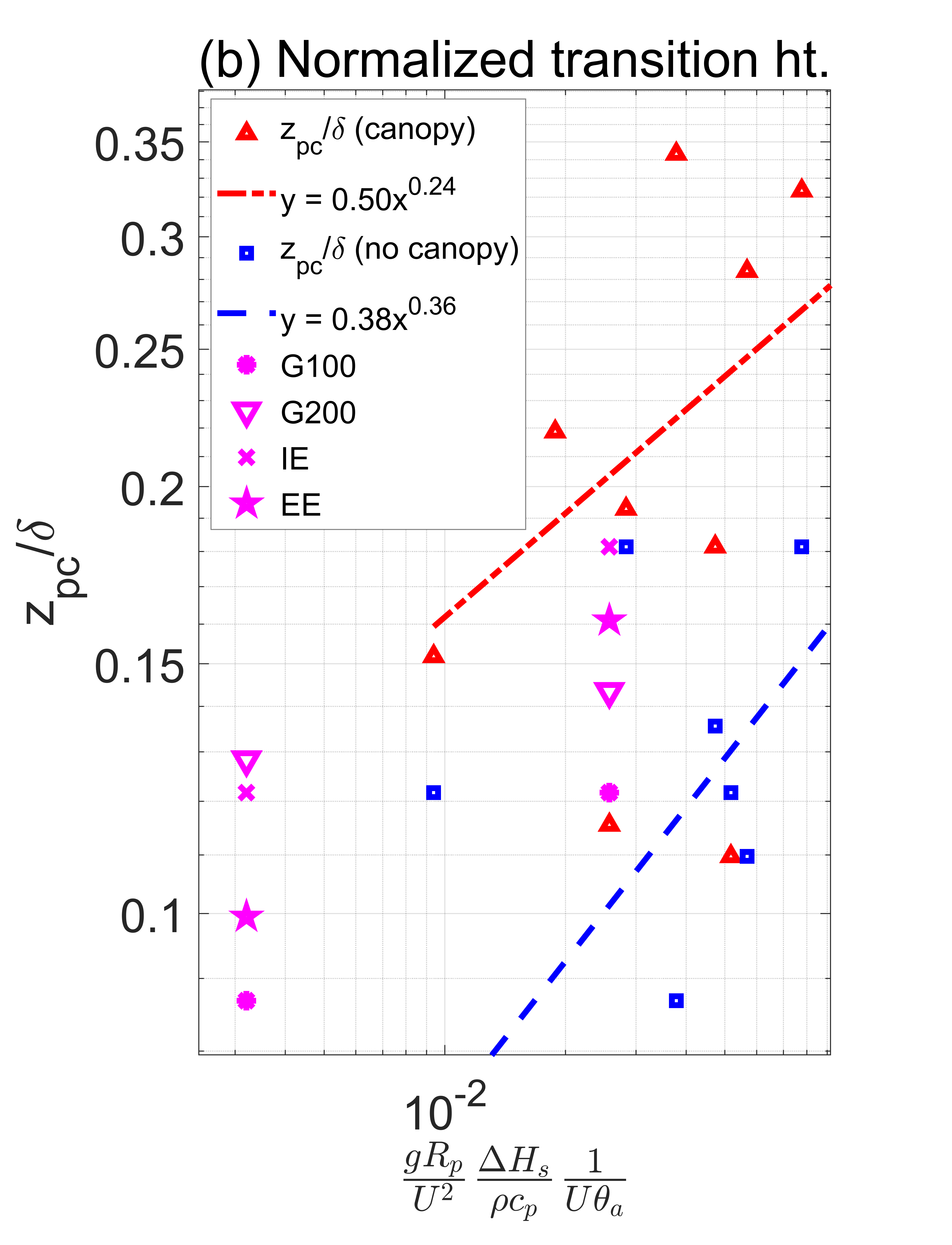}     
    \end{tabular}
    
    \caption{(a) Plot comparing the normalized height ($z_\text{pc}/\delta$) at which the plume transitions from the rise phase to the far-field bent-over phase (Zone 2) between the canopy and no-canopy cases. Trend-lines for $z_\text{pc}/\delta$ represent power-law dependence; $L_\text{B}/\delta$ represents the normalized buoyancy length scale. (b) Zoomed-in version of (a) with outliers from the heterogeneous-canopy cases simulated by Cervantes et al. \cite{cervantes2026impact} included }
    \label{fig_ztrans_nocan}
\end{figure}

\subsubsection{Canopy: Dependence on canopy height and density}
\label{sect_pi5pi6}
We now explore the impact of $\pi_6$ ($h_\text{c}/R_\text{p}$) and $\pi_7$ ($L_\text{c}/R_\text{p}$) on the plume centerline slopes and transition height into the bent-over phase, where $ L_\text{c} \approx h_\text{c}(c_\text{d}LAI)^{-1}$. The comparison between the canopy and no-canopy cases presented so far and in previous studies \cite{desai2025investigating, cervantes2026impact} could lead us to hypothesize that the near-surface centerline slope and transition height increase with increasing canopy height ($h_\text{c}$). An increase in the tree drag coefficient and the canopy density ($c_\text{d}LAI$) is also expected to result in the same outcome. However, canopy effects on a rising plume can be influenced by a complex interplay between canopy drag on plume updrafts, canopy drag on ambient horizontal momentum, and turbulent momentum fluxes near the canopy top, making them non-intuitive. 
While an increasingly dense canopy can impose a higher drag on the horizontal momentum near the surface, it can also attenuate vertical momentum near the canopy top and the exchange of momentum with the atmosphere aloft. This may result in similar turbulence characteristics for extremely sparse and extremely dense canopies \cite{chaudhari2026impact}, further impacting the near-surface plume centerline slope. Moreover, while tall and highly dense canopies are strong horizontal momentum sinks, they can also attenuate buoyant updrafts originating from the surface heat source. In contrast, short, highly sparse canopies impose a relatively weaker drag on both horizontal momentum and buoyant updrafts. Therefore, canopies at the extremes of the density spectrum may exhibit similar effects on the plume centerline slopes. Table~\ref{tab:hc_LAI_effects} summarizes the possible impacts of varying the canopy heights and densities and demonstrates the complexity of their combined effects on the near-surface plume and wind conditions. Additionally, it is difficult to characterize the effect of these canopy features on the far-field plume centerline.\\

\textbf{Plume centerline slopes.} To visualize the aforementioned complexity, 
we construct a dimensionless $\pi$ group ($\hat{\pi}_6$) by amalgamating $\pi_6$ and $\pi_7$ as follows:

\begin{equation}
    \hat{\pi}_6 = \frac{\pi_6^2}{\pi_7} = \left(\frac{h_\text{c}}{R_\text{p}}\right)^2\left(\frac{R_\text{p}}{L_\text{c}}\right) = \frac{h_\text{c}}{R_\text{p}}c_\text{d}LAI.
\end{equation}

\begin{table}[h!]
    \centering
    \begin{tabular}{p{35mm}|p{10mm}|p{15mm}|p{50mm}|p{35mm}}
     Canopy characteristics    & $h_\text{c}$ & $(c_\text{d}LAI)^{-1}$ & Expected effect on local, near-surface wind conditions & Limiting Case\\
     \hline
     (i) Tall, dense    & High & Low & Strong updraft attenuation, high horizontal momentum sink& Raised solid wall \\
    (ii) Short, dense & Low & Low & Updraft attenuation, horizontal momentum sink & Flat surface\\
    (iii) Short, sparse & Low & High &  Weak updraft attenuation, weak horizontal momentum sink& Flat surface\\
   (iv)  Tall, sparse & High & High & Weak updraft attenuation, horizontal momentum sink & High roughness length on flat surface
    \end{tabular}
    \caption{A summary of the expected effect on local, near-surface wind conditions of canopy height and drag/density}
    \label{tab:hc_LAI_effects}
\end{table}

\noindent Figure~\ref{fig_slopes_pi6pi7}(a) depicts the plume centerline slopes in Zones 1, 1(B), and 2 versus $\pi_6^2/\pi_7$ ($\hat{\pi}_6$). The advantage of choosing $\hat{\pi}_6$ lies in that its lowest and highest extremities represent the shortest, sparsest canopies and the tallest, densest canopies, respectively, aiding in our visualization. 
Figure~\ref{fig_slopes_pi6pi7}(a) reveals the relative order of magnitude of the plume centerline slopes in each zone, with Slope 1 being the highest and Slope 3 the lowest. Moreover, while the relative spread in Slope 2 appears to be low (coefficient of variation being 0.19), the relative spread in Slopes 1 and 3 is comparatively much higher, with coefficients of variation of 0.84 and 0.59, respectively. While Fig.~\ref{fig_slopes_pi6pi7}(a) helps obtain a general description of the centerline slopes, exploring a power-law scaling dependence on $\pi_6^2/\pi_7$ would impose an artificial restriction on the ratio between the exponents of $\pi_6$ and $\pi_7$. Therefore, we pursue a multivariate power-law regression, where the exponents of $\pi_6$ and $\pi_7$ are determined using a simplex optimization algorithm (MATLAB's \textit{fminsearch}()). The following power-law dependencies are revealed: Slope 1 $\sim\pi_6^{0.93}\pi_7^{0.13}$, Slope 1(B) $\sim\pi_6^{-0.18}\pi_7^{0.06}$, and Slope 2 $\sim\pi_6^{0.21}\pi_7^{0.04}$, which are presented in Figs.~\ref{fig_slopes_pi6pi7}(b)--(d) and can be recast as follows.
\begin{subequations}
    \begin{align}
\left(\frac{\Delta \overline{z}}{\Delta x}\right)_{1}\sim 5.50 \left[\left(\frac{h_\mathrm{c}}{R_\mathrm{p}}\right)^{7.26}\left(\frac{L_\mathrm{c}}{R_\mathrm{p}}\right)\right]^{0.13} = 5.50 \left(\frac{h_\mathrm{c}}{R_\mathrm{p}}\right)^{0.93}\left(\frac{L_\mathrm{c}}{R_\mathrm{p}}\right)^{0.13}  \equiv 5.50 \left(\frac{h_\mathrm{c}}{R_\mathrm{p}}\right)^{1.06}\left(c_\text{d}LAI\right)^{-0.13} \label{eq_slopes_pi6pi7_a}\\
\left(\frac{\Delta \overline{z}}{\Delta x}\right)_{\text{1(B)}}\sim 1.13 \left[\left(\frac{h_\mathrm{c}}{R_\mathrm{p}}\right)^{-2.83}\left(\frac{L_\mathrm{c}}{R_\mathrm{p}}\right)\right]^{0.06} = 1.13 \left(\frac{h_\mathrm{c}}{R_\mathrm{p}}\right)^{-0.18}\left(\frac{L_\mathrm{c}}{R_\mathrm{p}}\right)^{0.06}   \equiv 1.13 \left(\frac{h_\mathrm{c}}{R_\mathrm{p}}\right)^{-0.12}\left(c_\text{d}LAI\right)^{-0.06}\label{eq_slopes_pi6pi7_b}\\
\left(\frac{\Delta \overline{z}}{\Delta x}\right)_{\text{2}}\sim 0.19 \left[\left(\frac{h_\mathrm{c}}{R_\mathrm{p}}\right)^{4.82}\left(\frac{L_\mathrm{c}}{R_\mathrm{p}}\right)\right]^{0.04} = 0.19\left(\frac{h_\mathrm{c}}{R_\mathrm{p}}\right)^{0.21}\left(\frac{L_\mathrm{c}}{R_\mathrm{p}}\right)^{0.04}   \equiv 0.19\left(\frac{h_\mathrm{c}}{R_\mathrm{p}}\right)^{0.25}\left(c_\text{d}LAI\right)^{-0.04}
\label{eq_slopes_pi6pi7_c}
    \end{align}
    \label{eq_slopes_pi6pi7}
\end{subequations}
Near the surface, the plume centerline slope (Slope 1) increases as a function of both the dimensionless canopy height ($h_\text{c}/R_\text{p}$) and the dimensionless adjustment length scale ($L_\text{c}/R_\text{p}$). The inverse power-law relationship with $c_\text{d}LAI$ with a smaller exponent suggests that the near-surface slope decreases mildly with canopy drag, while the higher exponent for $h_\text{c}/R_\text{p}$ suggests that the increase in Slope 1 is steeper and almost linear with canopy height relative to the decrease with canopy drag (Eq.~\eqref{eq_slopes_pi6pi7_a}). This suggests that, for a given $c_\text{d}LAI$, increasing the canopy height is likely to weaken horizontal momentum more sharply than vertical buoyant updrafts, despite the longer adjustment (or drag) length scale, i.e. $L_\mathrm{c}$. In other words, a taller canopy, sharing the same $c_\text{d}LAI$ as a shorter canopy, is effectively ``sparser", resulting in a higher plume centerline slope. Equation~\eqref{eq_slopes_pi6pi7_a} also suggests that plumes arising from heat sources within the shortest canopies, whether sparse or dense, have centerlines that are more strongly deflected from the vertical. Therefore, plume centerline inclination angles in Zone 1 are similar for Cases (ii) and (iii) of Table~\ref{tab:hc_LAI_effects}, in the limiting cases.
\begin{figure}[h!]
    \centering
    \begin{tabular}{cc}
      \begin{tabular}{c}
         \includegraphics[width=0.45\textwidth, trim={3mm 0mm 2mm 3mm},                clip]{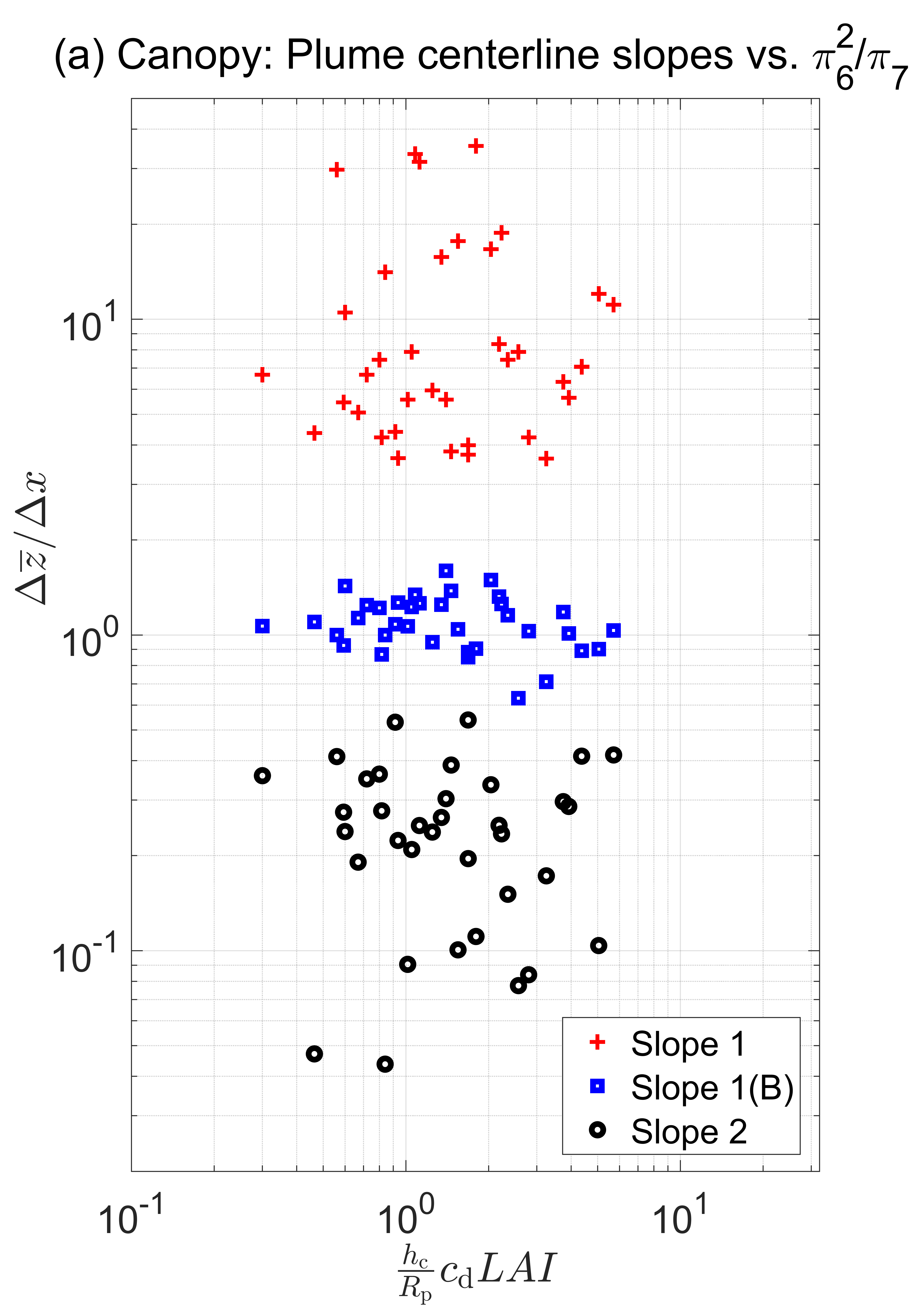}\\
         \includegraphics[width=0.46\textwidth]{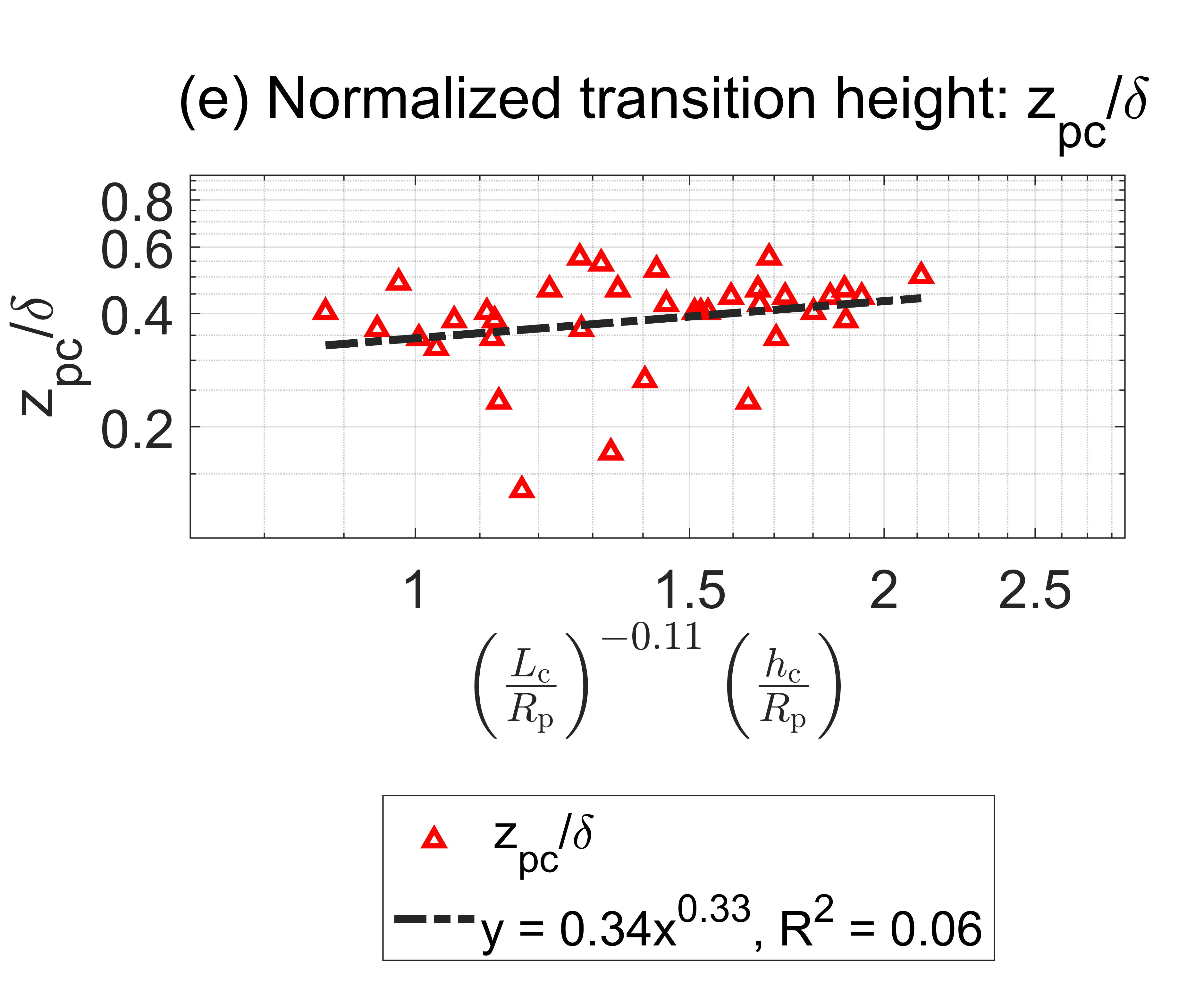}
       
      \end{tabular}
        &
        \begin{tabular}{c}
            \includegraphics[width=0.35\textwidth, trim={2mm 4mm 12mm 12mm},
                clip]{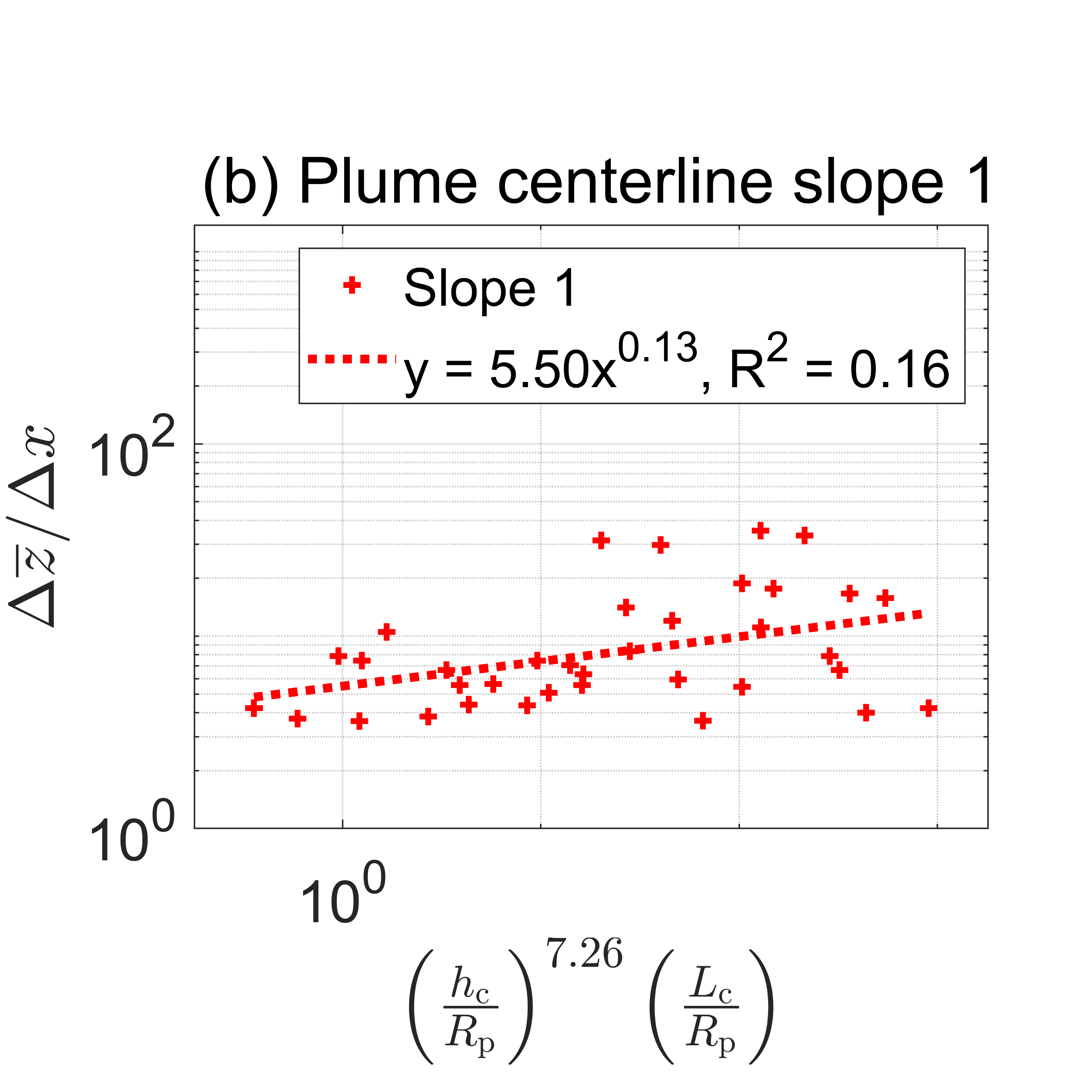} \\
             \includegraphics[width=0.35\textwidth, trim={2mm 4mm 7mm 12mm},
                clip]{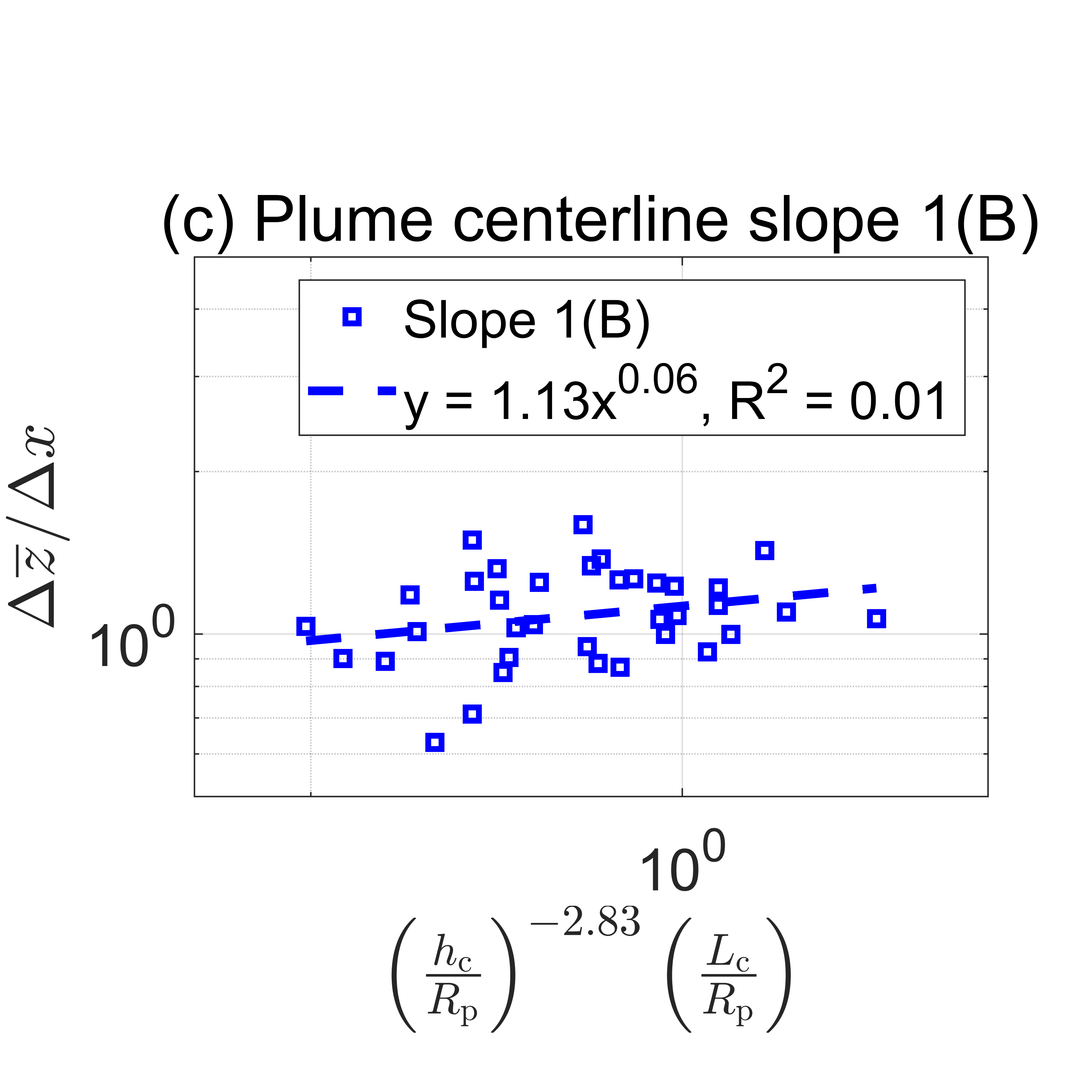} \\
            \includegraphics[width=0.35\textwidth, trim={2mm 2mm 10mm 12mm},
                clip]{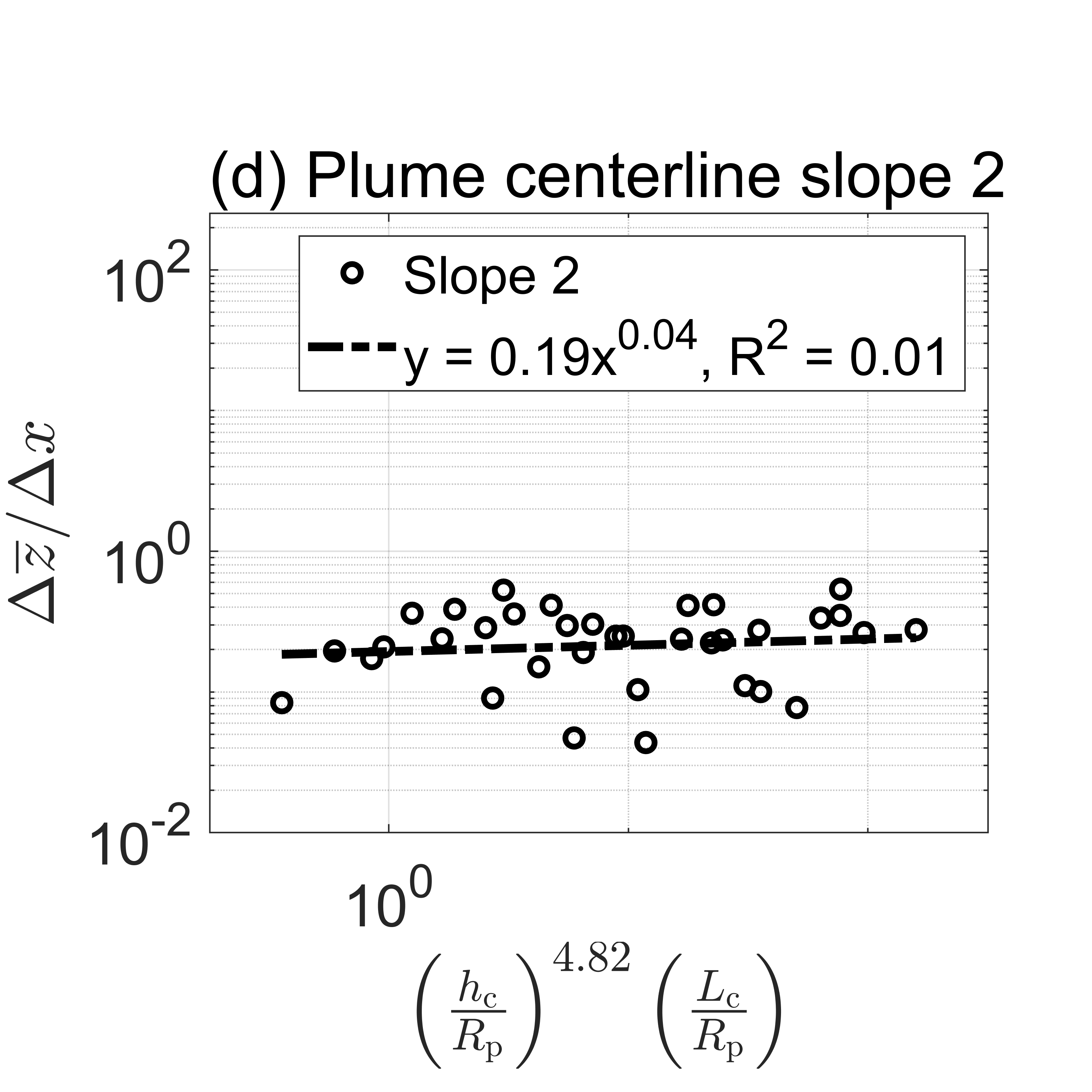}
        \end{tabular}
    \end{tabular}
    \caption{Plume centerline slopes in Zones 1, 1(B), and 2 plotted against (a)$\pi_6^2/\pi_7$. Slopes (b) 1, (c) 1(B), and (d) 2, and the (e) normalized transition height plotted against $\pi_6$ and $\pi_7$ with their respective power-law exponents on a log-log scale}
    \label{fig_slopes_pi6pi7}
\end{figure}

Realistically (and in the cases considered here), $c_\text{d}LAI$ is always non-zero, finite, and bounded. In the hypothetical limiting case, in which $c_\text{d}LAI$ would become very large, the dense canopy acts as a solid surface at the canopy top, decoupling the flow from the atmosphere aloft \cite{chaudhari2026impact}. This causes the plume centerline to deflect more steeply just above the canopy, resulting in a lower average centerline slope near the surface, akin to the no-canopy case. This is corroborated by Eq.~\eqref{eq_slopes_pi6pi7_a}. As $c_\text{d}LAI$ decreases (e.g., as trees shed their foliage), both buoyant updrafts and ambient horizontal momentum encounter less drag. However, Eq.~\eqref{eq_slopes_pi6pi7_a} suggests that the overall result is an increase in the near-surface plume centerline slope. Therefore, with reference to Table~\ref{tab:hc_LAI_effects}, Slope 1 in Case (i) is shown to be lower (more gradual) compared to Slope 1 in Case (iv).


Similarly, the plume centerline slope in the far field (Slope 2) is observed to be more sensitive to the canopy height compared to canopy drag, following a $1/4^\mathrm{th}$ power law with $h_\mathrm{c}/R_\mathrm{p}$ as shown by Eq.~\eqref{eq_slopes_pi6pi7_c}. As expected, the sensitivity of the plume slope to canopy height is higher near the surface compared to the far field. In the intermediate zone (Zone 1(B)), there appears to be an inverse power-law dependence on both the canopy height and drag.

 However, the power-law exponents suggest a much weaker dependence and lower sensitivity to these parameters compared to Zones 1 and 2. This is also in consonance with the low relative spread in Slope 1(B) as seen from Fig.~\ref{fig_slopes_pi6pi7}(a). \\

\textbf{Plume transition altitude.} Furthermore, we have presented power-law dependencies of the altitude at which the plume transitions to the far-field bent-over phase, normalized by the ABL height, on $\pi_6$ and $\pi_7$, in Fig.~\ref{fig_slopes_pi6pi7}(e). These dependencies can be recast as shown below: 

\begin{equation}
    \frac{z_\text{pc}}{\delta} \sim 
    0.34 \left[\left(\frac{L_\mathrm{c}}{R_\mathrm{p}}\right)^{-0.11}\left(\frac{h_\mathrm{c}}{R_\mathrm{p}}\right)\right]^{0.33} = 0.34 \left(\frac{L_\mathrm{c}}{R_\mathrm{p}}\right)^{-0.04}\left(\frac{h_\mathrm{c}}{R_\mathrm{p}}\right)^{0.33} \equiv 0.34 \left(c_\text{d}LAI\right)^{0.04}\left(\frac{h_\mathrm{c}}{R_\mathrm{p}}\right)^{0.29}
    \label{eq_ztrans_pi6pi7}
\end{equation}

Equation~\eqref{eq_ztrans_pi6pi7} confirms that an increase in either canopy drag or height results in an increase in the plume transition height. However, it is worth noting again that the plume transition height has a stronger power-law dependence and, thereby, a higher sensitivity to the dimensionless canopy height than to the dimensionless adjustment length scale or canopy drag. Thus, buoyant plumes originating from taller canopies are more likely to transition to the far-field bent-over phase at higher altitudes as opposed to denser canopies, potentially due to the combined, debilitating effect of the latter on both vertical as well as horizontal momentum, both within the canopy and in the region above the canopy top. It must be noted that these dynamics remain to be explored with a higher proportion of ``bottom-heavy"  canopies using the current computational setup.


\section{Conclusions and Future Work}
\label{sect_conclusions}
In this study, we have explored the dependence between the centerline slope of a buoyant plume emanating from a static source, akin to a surface fire, as it interacts with an ambient crosswind and a set of influencing variables, including the heat source strength, ambient wind speed, source geometry, and turbulence intensity, both in the presence and absence of a vegetative canopy. We have utilized the Buckingham Pi Theorem to reduce the influencing variables to a set of dimensionless parameters or $\pi$ groups, primarily describing the relative strength of buoyancy to the ambient wind forcing (similar to Byram's convective number or the Clark convective Froude number), the non-dimensionalized turbulence intensity, and canopy length scales relative to the source geometry. 
Following the dimensional analysis, we have demonstrated that the plume centerline slope can be obtained using scaling laws previously reported in the literature, primarily from laboratory experiments, in terms of the $\pi$ groups, thereby making them robust. Next, we have attempted to quantify the dependence of the plume centerline slope on the $\pi$ groups, using LES results from a suite of simulations with systematic variation in the ambient wind speed, heat source strength, and canopy height, density, and drag characteristics. Given the particular sparseness of scaling laws in the literature characterizing this dependence in canopy environments, we have attempted to obtain power-law dependencies from (a) a set of simulations conducted both in the presence and absence of an idealized, homogeneous, tall vegetative canopy with features resembling those of an Amazon rain forest, facilitating a comparison, and (b) a second set of simulations confined to the canopy setting, but with varying canopy geometry. 

The LES results from Set (a) suggest that plume centerline slopes near the surface are an order of magnitude higher in the canopy cases as compared to the no-canopy cases, owing to the drag afforded by the canopy to the rising plume. In contrast, far-field plume centerline slopes are of a similar order of magnitude in both cases. Increasing trends are observed as a function of $\pi_3\pi_4$ $\left(\frac{gR_\text{p}\mathrm{\Delta} H_\text{s}}{U^3\rho c_p \theta_\text{a}}\right)$, which encapsulates the relative strength of buoyancy to the ambient wind forcing, in both cases. Power-law exponents describing this trend are similar between the two scenarios for the near-surface plume centerline slopes. However, this exponent is higher for the far-field plume centerline in the canopy case, suggesting that the far-field plume trajectory is influenced by and is more sensitive to an increase in $\pi_3\pi_4$ in the presence of the canopy. While plume centerline slopes in the heterogeneous-canopy cases are found to deviate considerably from our power-law model they increase with $\pi_3\pi_4$, which is physically consistent. 

It is found that the near-surface plume slope has a much stronger $1/4^\mathrm{th}$ power-law dependence on, i.e. higher sensitivity to, the turbulence intensity on the leeward side of the plume relative to the turbulence intensity at an upstream location ($u_*^{(\text{pl})}/u_*^{(\text{up})}$), in canopy as compared to no-canopy environments. This is due to the difference in the spatial structure of the momentum fluxes within the plume between the two environments close to the plume source. As the plume turbulence intensity on the leeward side increases in the canopy case, so does the intensity of upward momentum transport, resulting in an increase in the near-surface plume slope. In contrast, in the no-canopy case, an increase in the leeward plume turbulence intensity represents an increase in drag-inducing downward momentum transport. In the far-field region, a higher sensitivity of the plume centerline slope to $u_*^{(\text{pl})}/u_*^{(\text{up})}$ is also noticeable in the canopy case compared to the no-canopy case, suggesting a far-reaching influence of the canopy-induced turbulent interactions with the plume. However, this difference in sensitivity to $u_*^{(\text{pl})}/u_*^{(\text{up})}$ is not as palpable as it is closer to the surface. In addition, the dependence of the plume centerline slopes on $u_*^{(\text{pl})}/u_*^{(\text{up})}$  is more complex in the heterogeneous-canopy cases as the behavior of the turbulent momentum fluxes, both upstream and at the leeward plume edge, deviate considerably from their homogeneous counterpart. This results in considerable deviations of these slopes from the power-law model obtained here.

Additionally, we have found that the height at which the rising plume transitions to the far-field bent-over phase increases with $\pi_3\pi_4$ through $1/3^\mathrm{rd}$ and $1/4^\mathrm{th}$ power law dependencies in the no-canopy and canopy environments, respectively. The transition height is taller in the canopy case, since the aerodynamic effects of the canopy serve to keep the plume upright and delay the onset of the bent-over phase. However, the effect of the canopy on the transition height becomes less prominent when the buoyancy strength is much higher compared to the ambient wind forcing. Again, the transition heights for the heterogeneous-canopy cases are observed to deviate from the power-law model. However, they are observed to increase with $\pi_3\pi_4$, similar to the homogeneous-canopy cases. This is physically consistent, since an increase in the buoyancy source strength relative to ambient wind forcing should result in higher plume transition altitudes for any given canopy structure. 

Our exploration of power-law dependencies between the plume centerline slopes and canopy geometric characteristics (Set (b)) has revealed stronger dependence or higher sensitivity to canopy height as opposed to canopy drag and density. Near the surface, the centerline slope is found to increase almost linearly with canopy height, while the slope in the far field follows an approximately $1/4^\mathrm{th}$ power law. A weak inverse power-law dependence on canopy drag and density is found to exist, with even lower sensitivity in the far field. Our scaling analysis shows that buoyant plumes, arising from surface heat sources in taller canopies, are less likely to be deflected from the vertical near the surface, compared to shorter canopies, within a range of realistic canopy densities. The lower sensitivity of the near-surface slope to canopy drag is likely due to the proportionate impact on both buoyant updrafts and horizontal momentum for a given change in the canopy drag. Additionally, plume centerline slopes in the intermediate zone, where buoyancy and ambient forcing are more comparable, appear to be intrinsically the least sensitive to both canopy height and drag.  Finally, the plume transition height to the far-field bent-over phase appears to be more sensitive to the canopy height than canopy drag, within the range of realistic canopy densities considered here.

Future work is needed to address the limitations of this analysis. 
There are discrepancies in power-law exponents between the best fits to our LES results and experiments. These can be addressed by running more simulations over a wider range of values in the dimensionless parametric space to get more accurate estimates of the power-law exponents. It may also be helpful to investigate outliers in the plume centerline slopes and the causes for their deviation from the observed trends. While the effects of model constraints such as the domain size 
or the location of the plume source relative to the inlet have been shown to be minimal for a single combination of heat-source strength and ambient wind speed in a previous study, their effects on the plume physics, especially at the edges of the dimensionless parametric space, may need to be tested. 
Furthermore, uncertainties in quantifying the plume centerline slopes associated with the grid resolution in the far field %
remain to be quantified. While we have explored the effect of bulk canopy geometric characteristics such as canopy density and height predominantly for ``top-heavy" canopies, the effect of varying $LAD$ profiles for the same $LAI$ needs to be investigated in more detail, along with the effect of distinctions in the horizontal and vertical projected area of the canopy elements. As illustrated above, heterogeneities in the canopy structure result in considerable deviations from the obtained power laws. This warrants a further investigation of the dependence of plume centerline slopes on a set of dimensionless parameters that encapsulate canopy heterogeneity across various levels. In addition to the cases we have simulated here ($>90$ including the heterogeneous-canopy cases) , a plethora of supplementary simulations are required to assess the sensitivity to variations in these conditions. Alternative scaling laws (e.g., polynomial) may need to be explored to account for the variance in the plume slopes currently unexplained by power-law scalings. Moreover, the authors aspire to extend the scaling analysis to characterize the plume centerline deflection and rate of spread for propagating firelines in both canopy and no-canopy environments, in stable as well as unstable atmospheric conditions. 

Notwithstanding the limitations of the present work, our analysis provides physical insights 
into how forcing mechanisms such as buoyancy strength, ambient wind speeds, and turbulent momentum fluxes, informed by the presence of a canopy, impact plume trajectories. Our analysis opens pathways for the development of future scaling laws on buoyant-plume--crosswind--canopy interactions, which has hitherto represented a major gap in the literature. Such scaling laws can inform the development of improved, more efficient operational models for managing smoke hazards across a multitude of forcing conditions, which can find applications in mitigating the effects of natural and human-made disasters. 




\section*{Appendix: Formulating the $\pi$ Groups}

To formulate the $\pi$ groups, we first return to the functional dependence of the response variable $\lambda_\text{pc}$ on the geometric constraints, forcing variables, and boundary conditions:
 \begin{equation*}
      \lambda_\text{pc} = \mathcal{F}(\delta,~g, \frac{\mathrm{\Delta} H_\text{s}}{\rho c_\text{p}},~u_*,~h_\text{c},~L_\text{c},~R_\text{p},~\theta_\text{a},~U).
 \end{equation*}

\begin{table}[h!]
    \centering
    \caption{Formulating the $\pi$ groups}
    \small
    \begin{tabular}{p{20mm}|p{39mm}|p{28mm}|p{10mm}}
    \hline\noalign{\smallskip}

    $\pi$ Form & Steps 2 and 3  & Step 4 & Final form\\
\noalign{\smallskip}\hline\noalign{\smallskip}
\multicolumn{4}{c}{\text{\textbf{Horizontal plume length scale: $\lambda_\text{pc}$}}}\\
\noalign{\smallskip}\hline\noalign{\smallskip}
     \[\pi_1=\]\[\lambda_\text{pc} R_\text{p}^a~ \theta_\text{a}^b U^c\]   & $[\pi_1]$\newline$ = (L)(L^a)(\mathrm{\Theta})^b (L T^{-1})^c$ \newline $= L^{1+a+c}\mathrm{\Theta}^b T^{- c}$ & $b = 0$, \newline $ - c = 0 \Rightarrow c=0$,
        \newline $1 + a + c = 0 \Rightarrow a = -1$ & \[\frac{\lambda_\text{pc}}{R_\text{p}}\]\\     
        \noalign{\smallskip}\hline\noalign{\smallskip}   
   \multicolumn{4}{c}{\text{\textbf{ABL height: $\delta$ }}}\\             \noalign{\smallskip}\hline\noalign{\smallskip} 
     \[\pi_2 =\]\[\delta R_\text{p}^a~ \theta_\text{a}^b U^c\] & 
Similar to $\pi_1$ above, & Similar to $\pi_1$ above, & \[\frac{\delta}{R_\text{p}}\] \\\noalign{\smallskip}\hline\noalign{\smallskip}
\multicolumn{4}{c}{\text{\textbf{Buoyancy: $g$}}}\\
\noalign{\smallskip}\hline\noalign{\smallskip}
\[\pi_3=\]\[ g R_\text{p}^a~ \theta_\text{a}^b U^c\]   & $[\pi_3]$\newline$ =(LT^{-2})(L^a)(\mathrm{\Theta})^b (L T^{-1})^c$ \newline $=L^{1+a+c}\mathrm{\Theta}^b T^{-2- c}$ & $b = 0$
        \newline$ -2 - c = 0 \Rightarrow c=-2$
        \newline $1 + a + c = 0  \Rightarrow a = 1$ & \[g\frac{R_\text{p}}{U^2}\]\\
\noalign{\smallskip}\hline\noalign{\smallskip}
\multicolumn{4}{c}{\text{\textbf{Difference in the kinematic heat flux: $\mathrm{\Delta} H_\text{s}$}}}\\
\noalign{\smallskip}\hline\noalign{\smallskip}
\[\pi_4 =\] \[\frac{\mathrm{\Delta} H_\text{s}}{\rho c_\text{p}}R_\text{p}^a~ \theta_\text{a}^b U^c\] & $[\pi_4]$ \newline $= (\mathrm{\Theta} LT^{-1})(L^a)(\mathrm{\Theta})^b (L T^{-1})^c$ \newline  $= L^{1+a+c}\mathrm{\Theta}^{1+b} T^{-1- c}$ & $1+b = 0 \Rightarrow b = -1$
        \newline $ -1 - c = 0 \Rightarrow c=-1$
        \newline $1 + a +c = 0 \Rightarrow a = 0$ &  \[\frac{\mathrm{\Delta} H_\text{s}}{\rho c_\text{p}}\frac{1}{\theta_\text{a}U}\] \\
\noalign{\smallskip}\hline\noalign{\smallskip}
\multicolumn{4}{c}{\text{\textbf{Friction velocity: $u_*$}}}\\
\noalign{\smallskip}\hline\noalign{\smallskip}
\[\pi_5 =\]\[ u_*~R_\text{p}^a~ \theta_\text{a}^b U^c\] & $[\pi_5]$\newline$ =( LT^{-1})(L^a)(\mathrm{\Theta})^b (L T^{-1})^c$\newline$ = L^{1+a+c}\mathrm{\Theta}^{b} T^{-1- c}$& $b = 0$ 
        \newline $ -1 - c = 0 \Rightarrow c=-1$
        \newline $1 + a +c = 0 \Rightarrow a = 0$& \[ \frac{u_*}{U}\]\\
        \noalign{\smallskip}\hline\noalign{\smallskip}
        \multicolumn{4}{c}{\text{\textbf{Canopy height: $h_\text{c}$}}}\\
        \noalign{\smallskip}\hline\noalign{\smallskip}
        \[\pi_6 =\] \[h_\text{c}R_\text{p}^a~ \theta_\text{a}^b U^c\]&

Similar to $\pi_1$ above, & Similar to $\pi_1$ above, & \[ \frac{h_\text{c}}{R_\text{p}}\]\\
\noalign{\smallskip}\hline\noalign{\smallskip}
        \multicolumn{4}{c}{\text{\textbf{Canopy drag length scale: $L_\text{c}$}}}\\
        \noalign{\smallskip}\hline\noalign{\smallskip}   
 \[\pi_7 =\]\[ L_\text{c}R_\text{p}^a~ \theta_\text{a}^b U^c\] &  Similar to $\pi_1$ above,      & Similar to $\pi_1$ above, & \[ \frac{L_\text{c}}{R_\text{p}}\]\\
\noalign{\smallskip}\hline    \end{tabular}
    \label{tab:pi}
\end{table}

 According to the argument presented in Sect.~\ref{sect_methods}, the number of dimensionless groups is: $M_\text{d}-N_\text{d}= 7$. 
Having chosen the repeating variables to be (i) $R_\text{p}$ [$L$], (ii) $U$ [$L,~T$], and (iii)  $\theta_\text{a}$ [$\mathrm{\mathrm{\mathrm{\Theta}}}$], each remaining variable is combined with the repeating variables to obtain the $\pi$ groups. For each remaining variable ($V_\mathrm{i}$), we follow the steps below to arrive at the dimensionless $\pi$ group in Table~\ref{tab:pi}: 

\begin{enumerate}
    \item Start with the preliminary form $\pi_\mathrm{i} =$ $V_\mathrm{i} \times R_\text{p}^a~ \theta_\text{a}^b U^c$,

    \item Substitute the dimensions of the physical quantities on the right hand side (RHS), expand, and group the like terms,

    \item Sum the exponents of each of the 3 fundamental dimensions ($L,~T,~ \mathrm{\Theta}$) to zero,

    \item Solve the resulting system of equations simultaneously, 

    \item Arrive at the final form.
\end{enumerate}

\newpage
\section*{Acknowledgments}
 Banerjee acknowledges the funding support from the University of California Office of the President (UCOP) grant LFR-20-653572 (UC Lab-Fees); the National Science Foundation (NSF) grants NSF-AGS-PDM-2146520 (CAREER), NSF-OISE-2114740 (AccelNet), NSF-CPS-2209695, NSF‐ECO‐CBET‐2318718, NSF-DMS-
2335847, and NSF-RISE-2536815; the United States Department of Agriculture (USDA) grant 2021-67022-35908 (NIFA); and a cost reimbursable agreement with the USDA Forest Service 20-CR-11242306-072. Desai acknowledges funding support from the Graduate Dean's Dissertation Fellowship, the Distinguished Public Impact Fellowship, and the Henry Samueli Endowed Fellowship at UC Irvine. Desai acknowledges that part of this work was performed under the auspices of the U.S. Department of Energy by Lawrence Livermore National Laboratory under Contract DE-AC52-07NA27344 and was supported by the LLNL-LDRD Program under Project No. 26-ERD-047: the Release Number is \textbf{LLNL-JRNL-2024559}.

\section*{Data Availability Statement}
We used the PALM model (Maronga et al. 2015 \cite{maronga2015parallelized}, Maronga et al. 2020 \cite{ maronga2020overview}), which can be downloaded from its official site: \url{https://palm.muk.uni-hannover.de/trac/wiki/doc/install}. The parameter file and the Python script to generate the NETCDF (static driver) file with canopy geometry and surface heat flux information, which were modified for our analysis, can be found at \url{https://palm.muk.uni-hannover.de/trac/wiki/doc/app/plant_canopy_parameters}. 

\bibliographystyle{ieeetr}
\bibliography{references}

\end{document}